# The Dawn of Alchemical Free-Energy Methods in Biomolecular Simulations

Daniele Macuglia[a], Giovanni Ciccotti[b], Benoît Roux[c,1]

[a]Department of History of Science, Technology and Medicine, Academy for Advanced Interdisciplinary Studies, Peking University, Beijing 100871, China

[b]Department of Physics, University of Rome "La Sapienza", 00185 Rome, Italy; IAC--CNR, Institute for the Application of Computing "M. Picone," National Research Council, 00185 Rome, Italy; School of Physics, University College Dublin, Belfield, Dublin 4, Ireland

[c]Department of Biochemistry and Molecular Biology, University of Chicago, Chicago, IL 60637, USA.

[1]Email: roux@uchicago.edu

Abstract. From the onset of fundamental statistical mechanical constructs formulated in the late 19th century, alchemical free-energy methods slowly emerged and transitioned to become operational tools of biomolecular simulation applicable to a wide range of problems including protein–ligand binding for drug discovery research. This article reconstructs how statistical mechanical approaches such as thermodynamic integration and free-energy perturbation were reconfigured in the early 1980's to address the complexities of increasingly heterogeneous biomolecular systems. Drawing on oral history interviews and primary literature, the study examines the technical, institutional, theoretical, and infrastructural conditions under which these methods were implemented, and became progressively operational. These conditions encompassed the consolidation of lab-specific software infrastructures, the formulation of practical simulation protocols, as well as essential statistical mechanical clarifications. From this perspective, the progress of free-energy methods proceeded less from a unified convergence than from an iterative troubleshooting process of alignment involving practical and theoretical considerations. The aim of the present article is to offer a historically grounded account of how free-energy techniques acquired practical and functional reliability.

## 1. INTRODUCTION

The development of alchemical free energy computations represents a remarkable example of how fundamental knowledge is transformed and adapted to become a useable tool with real-world impact. Starting from fundamental statistical mechanical constructs formulated in the late 19th century,[1-6] alchemical free-energy methods slowly emerged and transitioned to become embedded instruments of biomolecular simulation applicable to a wide range of problems including protein–ligand binding contexts ultimately relevant to drug discovery.[7] With the advent of computer simulations emerging from the technological advances after the second world war in the 1950's,[8,9] formal theoretical constructs from statistical mechanics began to be applied to atomic and molecular systems of increasingly complexity. As confidence in these approaches matured with the advent of more realistic MD simulations of liquids,[10,11] researchers began experimenting with reweighting techniques that allowed the determination of thermodynamic information for different states near and beyond the supercritical point from a single ensemble.[12-15] Ultimately, multistage sampling techniques were introduced to enhance the precision of free-energy calculations particularly for systems characterized by complex interactions.[16,17] In 1976, Charles Bennett (born 1943) introduced a statistically optimal estimator for free-energy differences by leveraging overlapping ensembles and minimizing variance, now known as the Bennett Acceptance Ratio (BAR) method.[18] It is around this period that the first MD simulation of a folded protein was published.[19] This achievement was profoundly transformative. As people realized that MD simulations could be used as a sampling tool to "perform" statistical mechanics on complex biomolecular systems,[20] free-energy methodologies were rapidly integrated within a few years.

Suddenly, a few key studies played a major role in translating theoretical frameworks into biologically meaningful applications,[21,22] repositioning free-energy methods as potentially viable for problems in protein function, drug discovery, and molecular recognition. This was followed by a long period during which free-energy methods transitioned from abstract theoretical models to routine components of biomolecular simulation practice. Ultimately, the integration of these computational techniques within computer-aided drug design (CADD) efforts stands as a demonstration of their practical significance in industrial applications.[7,23] This achievement, nearly three decades after the earliest theoretical proposals is indicative of the operational maturity of free-energy computations. While improvements in computational capacity and algorithmic design played a role, the broader adoption of these techniques stemmed from multiple factors.

Drawing on oral histories and archival literature, we will examine here how these tools were made operational under conditions of institutional diversity, incomplete documentation, and variable user expertise. Rather than tracing a seamless path of algorithmic refinement, we highlight the distributed labor of software developments and all its challenges, optimization of force field parameters, and alignment with evolving modeling norms—processes often opaque in published accounts. The eventual dissemination of free-energy methods to drug discovery in the industry hinged on their ability to satisfy regulatory constraints, respond to modeling deadline constraints, and accommodate available computational resources. In this sense, reliability was not inherited from formal derivations alone, but was pieced together through ongoing technical mediation, social coordination, and persistent infrastructural adjustment.

Our fundamental objective is to reconstruct the thought process and the material and sociological circumstances of a diverse community of scientists whose work established the paradigm of biomolecular free-energy methodologies up to their application in modern-day drug development. Our aim, through this historical reconstruction (that is unavoidably colored by our own personal experience), is to highlight and better understand the origin of tentative ideas, the technical challenges of early implementations, the different decisions made at various crossroads, the episodic confusion and misunderstandings, and the occurrence of creative inspiration. For an exhaustive record of the field over the years, several excellent academic reviews are recommended.[24-31] It is our hope that the present document will clarify how scientific progress occurred, and help achieve a more profound understanding of the field as it is today.

## 2. HISTORICAL FOUNDATIONS OF FREE-ENERGY CALCULATION METHODS

The foundational works of Hermann von Helmholtz (1821–1894) and Josiah Willard Gibbs (1839–1903) in the late 19th century,[1-5] provided the mathematical and conceptual basis for the development of free-energy calculations in a wide range of applications. The use of free-energy methodologies spans from modeling bulk thermodynamic properties in idealized systems to simulating biologically relevant molecular interactions in complex environments. Initially framed within the context of classical thermodynamics, these methods were formalized through the definition of various thermodynamic potentials, such as Helmholtz ($A$) and Gibbs free energies ($G$), obtained using Legendre transformations to express these potentials in terms of natural variables such as volume or pressure, and entropy or temperature. In the context of statistical mechanics, the Helmholtz free energy is associated with the canonical distribution, while the Gibbs free energy arises naturally in the isothermal–isobaric regime, which is suited to systems maintained at constant pressure, particularly in experimental and biological settings. Helmholtz's contributions primarily focused on energy conservation and transformations under constant volume conditions, while Gibbs' framework extended to chemical equilibria and open-system behavior under pressure regulation. Gibbs' statistical mechanics formalism, in particular, systematized the use of ensemble averages and the partition function—quantities that linked microscopic phase space to macroscopic thermodynamic observables.

In generalizing the concept of free energy to encompass various ensembles (e.g., microcanonical, canonical, grand canonical), Gibbs laid the groundwork for evaluating thermodynamic potentials through the partition function, defined as the sum of the statistical weights of all accessible microstates in a given ensemble. This formulation provided a systematic means to calculate free energies and entropy based on the probabilistic distribution of microstates and offered a statistical mechanical interpretation of equilibrium, phase behavior, and chemical potentials.

Once the statistical framework was established, accounting for the permutation of identical particles, it enabled a rigorous definition of free energy by linking macroscopic thermodynamic properties to the microscopic configurations of a system. This formalism allowed for the derivation of key thermodynamic potentials, such as the Helmholtz free energy, $A = -k_B T \ln Z$, where $k_B$ is Boltzmann's constant, $T$ the temperature, and $Z$ the partition function. A similar derivation holds for the Gibbs free energy, each potential being anchored in its respective statistical ensemble. Gibbs' formulation thus established a comprehensive mathematical framework that connected thermodynamic potentials with statistical descriptions and laid a robust foundation for analyzing closed equilibrium systems. Yet, the practical integration of this formalism into computational workflows would unfold only decades later, shaped by evolving institutional capacities and the gradual articulation of simulation-based research programs.

As free-energy calculations developed greater sophistication in the early 20th century, following the foundational contributions of Gibbs, researchers confronted increasingly complex questions involving thermodynamic differences in systems that were not necessarily amenable to direct experimental measurement. This shift marked a broader transformation in statistical mechanics, moving away from analytically solvable models and toward the study of systems composed of interacting particles with complex couplings and solvent effects. These challenges called for theoretical and computational tools capable of bridging abstract formalism with measurable observables without requiring explicit enumeration of all microscopic states.

One of the earliest frameworks to address such problems was formulated by Max Born (1882–1970), whose theory of ion solvation[32] offered a physically intuitive model of solvation by treating the ion as a charged sphere embedded in a continuous dielectric medium. Born's formulation enabled the calculation of the reversible work required to charge the ion from zero to its full charge q in a medium of permittivity ϵ:

$$\Delta G = \int_0^Q \langle \phi(Q') \rangle dQ' \qquad \text{(Eq. 1)}$$

Here, $\langle \varphi(Q') \rangle$ denotes the average electrostatic potential experienced by a charge increment $dQ'$ in the dielectric medium. The equation partitions the total charging process into infinitesimal contributions of thermodynamic work, with each term representing the product of the local potential and the corresponding charge increment. Born's framework thus provided a basis for calculating the electrostatic free energy associated with charging an ion by integrating the potential at each incremental step up to the full charge $Q$, eq 1 standing as an early precursor of thermodynamic integration (TI). Although the specific formalism embodied by eq 1 is not found in the original text, it represents a contemporary re-interpretation that remains consistent with Born's underlying physical insights. The charging free energy expressed in terms of eq 1 has been widely adopted in modern accounts of the Born model for its clarity and practical applicability.[33] Peter Debye (1884–1966) and Erich Hückel (1896–1980) extended Born's model to ionic solutions, developing a mean-field continuum theory that captured the influence of ionic atmospheres on both potential and energy distributions.[34] Their approach introduced the concept of ionic strength—a parameter quantifying both the concentration and valence of ions in solution—which governs the effective screening of electrostatic interactions. This model enabled more realistic treatments of long-range interactions by incorporating the collective effects of mobile ions on the local electrostatic environment. While their work primarily addressed electrostatic potentials rather than free-energy differences per se, it established the conceptual link between microscopic interactions and macroscopic thermodynamic observables. In particular, their framework formalized the notion that free energy could be modulated by varying conjugate variables such as charge and electrostatic potential, further laying groundwork for TI in continuity with Born's treatment. Moreover, the Debye-Hückel theory's treatment of collective ionic behavior introduced a scalable framework for mean-field effects in condensed phases, a framework that proved essential for subsequent refinements of free-energy perturbation (FEP) and related techniques, especially in high ionic strength or long-range interaction scenarios.

By the mid-1930s, John Kirkwood (**Figure 1**) introduced a rigorous theoretical framework for TI to compute free-energy changes between two macroscopic states by integrating the derivative of the system's potential energy with respect to a dimensionless coupling parameter, λ, which continuously interpolates between initial and final Hamiltonians.[35] This formulation enabled controlled transformations in which the molecular interactions, rather than external macroscopic variables, were gradually modified. Since the intermediate states generated along this path do not correspond to physical realizations, the method shall be termed "alchemical"[i] to highlight the mathematical transmutation of one atom or group into another via nonphysical intermediates.[21,22] Although earlier TI approaches based on natural macroscopic state variables such as temperature or pressure had been explored[9,41,42] (for this reason, we shall refer to the latter as "natural" TI). Kirkwood's formalism introduced a fundamentally different paradigm. Rather than integrating along equilibrium paths defined by external conditions, Kirkwood TI systematically perturbed the interaction potential, which enabled the calculation of free-energy differences between chemically distinct species. Despite its conceptual elegance, the application of Kirkwood's approach was at first limited by both the mathematical complexity of many-body terms and the lack of computational tools for high-dimensional numerical integration. As a result, the method remained largely within theoretical analysis until the development of molecular simulation techniques and the advent of high-performance computing infrastructure in the latter 20th century.[20]

While absolute free energies quantify the total thermodynamic potential of a single state or system, they are highly sensitive to the choice of reference and typically require very extensive—and often impractical—sampling of the entire configuration space (with the momentum component of phase space usually separable and analytically integrable in classical mechanics). In contrast, free-energy differences measure relative changes between two defined macrostates or systems, naturally capturing the thermodynamic work associated with transformations (e.g., phase changes, chemical reactions, or conformational transitions). Kirkwood's original formulation[35]—later reframed through the lens of alchemical transformations—introduced a dimensionless coupling parameter "λ" to continuously interpolate between initial and final states along a well-defined thermodynamic path.[ii]

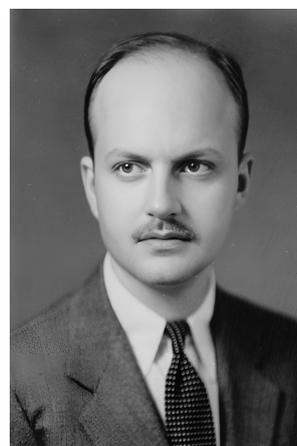

**Figure 1.** John Gamble Kirkwood (1907–1959), theoretical chemist whose work in statistical mechanics and the theory of liquids established formalisms central to molecular simulation, including the Kirkwood coupling-parameter integration method for free-energy differences. His research shaped the conceptual framework that underpins modern alchemical methodologies. Photograph dated 1938 (Smithsonian Institution Archives, Accession 90-105, Science Service Records, Image No. SIA2008-4847).

By the late 1980s, Kirkwood TI had been generalized to a variety of such transformations, including amino acid mutations in proteins and the systematic alteration of ligand chemical identities.[22] These applications repositioned the method within increasingly complex molecular contexts and allowed for free-energy calculations in biological systems and chemically diverse environments. The free-energy difference, $\Delta G$, is computed by integrating the statistical average of the derivative of the potential energy with respect to λ:

$$\Delta G = \int_0^1 \langle \frac{\partial U(\lambda)}{\partial \lambda} \rangle_\lambda d\lambda \qquad \text{Eq. (2)}$$

where $U(\lambda)$ represents the potential energy as a function of the coupling parameter; λ = 0 corresponds to the initial system with potential energy $U_0$, and λ = 1 to the final system $U_1$. The angle brackets $\langle \ldots \rangle_\lambda$ denote ensemble averages taken over the equilibrium distribution defined by $U(\lambda)$ at a specific value λ.[iii] In practice, the integral in Eq. (2) must be calculated by numerical quadrature from the discrete set of λ values sampled in the simulation. By analogy with the potential of mean force, Eq. (2) can be understood as the reversible work done by the mean force $\langle \partial U/\partial \lambda \rangle_\lambda$ along the generalized "coordinate" λ.

This formalism established a generalized mathematical basis for computing free-energy differences between states not directly accessible to experiment. Its generality enabled a range of practical implementations as simulation techniques matured. In emphasizing relative (rather than absolute) free energies, Kirkwood TI became a key tool for estimating thermodynamic changes under well-controlled conditions and laid the conceptual foundation for modern alchemical free-energy methods. The ability of the Kirkwood framework to map continuous transformations between chemically distinct species, without requiring physically realizable intermediates, proved especially valuable in pharmaceutical science, where predictive modeling of binding affinities and selectivity is crucial. To this day, Kirkwood TI remains integral to the computational design and refinement of molecular interactions.

In 1954, Robert Zwanzig (**Figure 2**) introduced the free-energy perturbation (FEP) method,[46] which provided a rigorous statistical framework for evaluating free-energy changes between two distinct macroscopic states of the same system.[iv] In contrast with Kirkwood's TI, which computes the same difference through an integration over a continuous transformation pathway, Zwanzig's FEP method is based on a statistical reweighting of configurations from a reference state, where the free energy difference is computed as an ensemble average of the exponential of the energy difference between the reference and target states.[46] The FEP equation is expressed as:

$$\Delta G = -k_\mathrm{B} T \ln \langle e^{-\Delta U / k_\mathrm{B} T} \rangle_0 \quad \text{(Eq. 3)}$$

where $\Delta U$ represents the energy difference between the two states (or systems), $T$ is the absolute temperature, and the angle brackets $\langle \cdots \rangle_0$ denote an ensemble average taken over the equilibrium distribution of the reference state, characterized by the potential energy $U_0$. Unlike TI, which requires evaluating ensemble averages along a continuous parametric pathway (typically governed by a coupling parameter $\lambda$), FEP evaluates a single ensemble average to estimate the free-energy difference from a finite perturbation.[47,48]

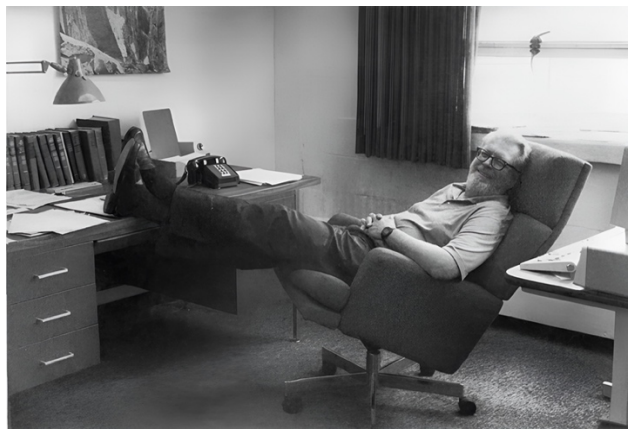

**Figure 2.** Robert W. Zwanzig (1928–2014), physical chemist at the University of Maryland and later at NIST. He formulated the 1954 statistical-mechanical perturbation theory (Zwanzig equation), which became the formal foundation for modern alchemical free-energy calculations. Photographed in his office at the Institute for Physical Science and Technology (IPST), University of Maryland, in the late 1980s. Photo by Al Danegger. Reprinted (adapted) with permission from D. Thirumalai, S. C. Greer, Zwanzig dedication, J. Phys. Chem. 1989, 93 (19), 6883–6884. Copyright 1989 American Chemical Society." Resolution improved via AI treatment.

FEP can be applied in both alchemical and non-alchemical contexts. In alchemical FEP, the method is used to compute free-energy differences between chemically distinct systems, where the Hamiltonian is interpolated between two end states. This enables the calculation of transformations that are not physically realizable, such as mutating one amino acid residue into another. Conversely, in non-alchemical FEP, the perturbation is applied within the same system, and the free-energy change reflects small physical shifts, such as changes in temperature, pressure, or applied field. However, because the method relies on exponential reweighting, FEP performs best when the energy distributions of the initial and final states exhibit substantial overlap, ensuring that the contribution of each configuration is adequately sampled and that statistical errors remain controlled. Where TI assumes a controllable continuity across a parameterized transformation, FEP presumes statistical substitutability between ensembles, a distinction with major implications for implementation fidelity. When overlap is limited, reweighting becomes unstable and error prone. Kirkwood TI and Zwanzig FEP both derive from the same statistical mechanical principles, but differ in operational implementation, sensitivity to sampling, and practical convergence. Their mathematical equivalence holds only under idealized conditions, such as near-complete sampling or infinitesimal perturbations, but in real-world applications, each method presents distinct trade-offs. Both frameworks stand in marked contrast to the original Gibbsian formulations, which required direct enumeration or summation over all microstates—a task intractable for even moderately complex systems. The ensemble averages in Eq. (2) and Eq. (3) yield reliable estimates under conditions of ergodicity, where observables such as energy differences (in FEP) or energy derivatives (in TI) vary smoothly across sampled trajectories. As sampling increases, statistical fluctuations diminish, improving convergence.[v] Both TI and FEP have their own advantages and tradeoffs; TI can remain stable even with limited ensemble overlap, provided the integrand varies smoothly, while FEP is highly efficient when overlap is strong but deteriorates rapidly when distributions separate.

Following the foundational contributions of Kirkwood and Zwanzig, Benjamin Widom (1927–2025) introduced in 1963 a conceptually distinct approach that further enriched the statistical treatment of free-energy changes.[53] Known as the "Widom insertion method," this technique estimates the chemical potential of a fluid by statistically evaluating the energy cost associated with the hypothetical insertion of a test particle into the system. Because the insertion process is evaluated without physically altering the system, the method allows one to infer the excess chemical potential from ensemble averages of Boltzmann factors associated with the inserted particle's potential energy.[vi] Widom's method, which infers chemical potentials via virtual test-particle insertions, proved especially effective in low-density systems, where steric interference is rare and successful insertion events more probable. Its relevance lies in the conceptual bridge it forms with FEP: both approaches use ensemble-based reweighting to estimate free-energy differences, though the Widom method is not alchemical in nature. Its performance degrades sharply in dense or structured systems (e.g., biological macromolecules) where successful insertions are rare and the statistical signal becomes dominated by rare events. For this reason, Widom's approach has been primarily applied to idealized systems and theoretical models, where it remains an elegant tool for exploring the thermodynamic implications of particle addition.

## 3. FROM CONDENSED MATTER PHYSICS TO BIOLOGICAL SIMULATIONS

The following years were marked by a formative period of methodological creativity where simulation techniques were deployed to calculate free-energy differences. With the growing use of computer simulations, the Monte Carlo (MC) method, originally developed by Metropolis et al.,[8] and the classical molecular dynamics (MD) from Alder and Wainwright,[9] began to be widely applied to more complex situations. As access to computing resources expanded, researchers began experimenting with reweighting techniques to bypass some of the limitations associated with absolute free-energy computations. Rather than representing a linear progression, these methodological shifts reflected a growing repertoire of statistical tools calibrated to specific physical models and computational constraints. Two closely linked studies by Ian McDonald (1938–2020) and Konrad Singer (1917–2013),[12,13] both affiliated with Royal Holloway College (University of London), illustrate how MC methods were adapted and repurposed in response to these shifting constraints. In their 1967 paper,[12] McDonald and Singer applied the Metropolis MC algorithm to simulations of liquid argon using Lennard-Jones potentials. They examined system behavior across varying volumes and temperatures and, drawing on earlier work by William Wood (1924–2005) and Frederick Parker,[14] introduced a reweighting technique that allowed a single sampling of the ensemble to yield thermodynamic information at multiple temperatures. This statistical reinterpretation of a single sample of the data was not designed to produce absolute free energies directly but instead enabled efficient estimation of thermodynamic observables across adjacent temperatures. They derived the relation:

$$G_n(T') = G_n(T) \cdot e^{-\phi_n(\beta-\beta')} \quad (Eq. 4)$$

Here, $G_n(T)$ denotes the statistical weight of configuration $n$ at temperature $T$, and $G_n(T')$ is the reweighted probability at $T'$, with $\phi_n$ representing the potential energy of the configuration, and $\beta - \beta' = 1/k_B T - 1/k_B T'$. This approach permitted the estimation of energy-dependent probability densities at new temperatures without re-sampling, which significantly enhanced computational efficiency and broadened the interpretive scope of simulation-based inference. In a follow-up study,[13] McDonald and Singer expanded this methodology to systems near and beyond the supercritical point. Incorporating parameters from Michels et al.,[54,55] they investigated thermodynamic quantities (e.g., the virial coefficient, excess entropy, specific heat) under conditions where experimental data were often unavailable or imprecise. The adaptability of their reweighting approach enabled simulations to probe regions characterized by sharp fluctuations and poor sampling convergence, which were traditionally inaccessible to conventional MC methods. These contributions exemplify how simulation practice became entangled with changing assumptions about model tuning, algorithmic feasibility, and physical representation. McDonald and Singer's work reflects an emergent pragmatics in simulation research, where algorithmic tools and thermodynamic assumptions were reconfigured to meet new theoretical and technical demands.

Concurrently, Jean-Pierre Hansen (born 1942) and Loup Verlet (1931–2019), from the *Laboratoire de Physique des Hautes Énergies* in Orsay, used MC simulations to study phase transitions in Lennard-Jones fluids.[15] Their study developed a method for estimating the Helmholtz free energy by integrating pressure–volume curves along isotherms, using constrained simulations to approximate homogeneous behavior across the gas-liquid coexistence region. This "equation-of-state" approach allowed them to analyze gas-liquid coexistence curves while avoiding the complexities of phase separation. To enforce homogeneity, they subdivided the system into cells and introduced bounds on local particle number fluctuations, effectively suppressing large-scale density fluctuations. This constraint-based sampling method produced smooth van der Waals-like isotherms, from which the Helmholtz free energy and phase coexistence points could be extracted using Maxwell constructions. Their methodology provided a systematic alternative to direct coexistence simulations and extended the utility of MC simulations for studying fluid phase equilibria. It also facilitated more accurate comparison with experimental data (e.g., argon) by reproducing the gas-liquid transition with greater thermodynamic fidelity. This constrained sampling strategy, though approximate, laid conceptual and computational groundwork for later free-energy methods by demonstrating how control over phase space sampling could yield robust thermodynamic predictions without needing to simulate explicit two-phase systems.[vii]

In 1972, Valleau (1932–2020) and Damon Card (1941–2014) at the University of Toronto introduced a multistage sampling technique to enhance the precision of free-energy calculations particularly for systems characterized by complex interactions, such as dipolar fluids.[16] Their work tackled foundational issues in ensemble sampling by segmenting the integration pathway into multiple overlapping distributions, a technique that mitigated the challenges posed by sharp energy discontinuities and poorly sampled rare events. Indeed, their approach provided a structural solution to the limitations of conventional MC methods, particularly the difficulty in sampling low-probability configurations essential for computing entropy and free energies. Instead of relying on a single configuration distribution, Valleau and Card constructed a series of intermediate "bridging" distributions, each overlapping with its neighbors and designed to interpolate between easily sampled reference states and high-energy target distributions. This design made it possible to evaluate the canonical partition function indirectly by anchoring it to known values at one end and extending it stepwise across the relevant energy spectrum. In contrast to natural TI, which requires pressure data over a continuous range of states, multistage sampling enabled a stepwise traversal of configuration space that proved especially effective for systems with strong repulsions or sharp interaction discontinuities. Applied to a two-component ionic fluid of hard spheres with Coulomb interactions, their method demonstrated accuracy comparable to existing approaches such as thermodynamic density integration, but with fewer simulations and improved control over convergence. In circumventing the need to sample rarely visited high-energy configurations directly, the multistage strategy provided a more flexible and computationally economical pathway for calculating excess free energy and entropy, especially in systems where direct estimation of the configuration integral is computationally prohibitive. The application of multistage sampling techniques was further expanded by Gren Patey (born 1948) and Valleau, who adapted the method to fluids with embedded point dipoles.[17] These systems presented novel challenges, as the anisotropic dipole–dipole forces introduced complex configurational landscapes requiring enhanced sampling strategies. Unlike isotropic fluids, where interactions are uniform in all directions, dipolar systems necessitated more careful control over the exploration of phase space to account for the directional and long-range nature of interactions. Patey and Valleau's refinement of the multistage approach enabled a more effective navigation of these energetic complexities, producing results that were systematically compared against mean spherical and perturbation theory models. Their comparison underscored the method's accuracy and scalability, especially under dense conditions. The 1974 study of Patey and Valleau addressed the inefficiencies associated with conventional Boltzmann-weighted sampling in systems with significant barriers or metastable states, particularly in systems near phase coexistence, where key configurations often lie in low-probability regions of phase space.[57] Employing a biased distribution tailored to emphasize regions of high relevance to the free-energy difference, importance sampling improved convergence and reduced variance in MC estimates. Rather than sampling the entire configuration space uniformly, their method concentrated effort on configurations that contribute meaningfully to the target observable, especially in systems characterized by steep energy gradients or rare event dynamics.

This redirection of computational strategies exemplifies the broader trend during the 1970s of reconfiguring statistical sampling practices to better match the physical features and computational constraints of complex fluid systems. These developments addressed key sampling bottlenecks and improved the feasibility of free-energy estimation in multidimensional configuration spaces.[50] Nonetheless, their application remained mostly confined to simple liquids or crystalline systems, with biological contexts still beyond reach. A major methodological shift arrived with the development of the umbrella sampling technique by Torrie (born 1949) and Valleau.[49,50] The technique was designed to overcome the inherent sampling inefficiencies of MC simulations in systems where rare events or phase transitions dominated the thermodynamic landscape.[49,50] Conventional MC methods, which sample configurations according to Boltzmann-weighted distributions, tend to neglect low-probability states—often the configurations most relevant for accurate free-energy estimates in such systems. This limitation becomes especially severe near phase boundaries or in systems characterized by sharp energy barriers. Umbrella sampling addressed this challenge by applying a biasing potential designed to flatten high-energy barriers, thereby facilitating enhanced sampling across the entire configuration space. The biasing function was tailored to favor configurations that contribute most significantly to the free-energy integrand. Rather than distorting the thermodynamic results, this deliberate bias was statistically corrected afterward by reweighting the sampled distributions, allowing researchers to recover unbiased thermodynamic quantities from biased ensembles. Torrie and Valleau tested this approach on Lennard-Jones fluids, including regions near the gas-liquid coexistence curve, where conventional methods struggled. Employing successive overlapping umbrella distributions (instead of a single, wide-ranging one), they ensured convergence even in regions characterized by steep energy gradients. These overlapping windows enabled accurate estimation of probability densities as small as $10^{-8}$, a significant computational achievement at the time. The technique proved especially effective when applied to "soft-sphere" to Lennard-Jones transformations and later to temperature-scaling pathways, which enabled estimation of Helmholtz free energy across broad thermodynamic ranges from only a few simulations.[viii]

Around the same time, the conceptual landscape of FEP methods was expanded with the introduction by Charles Bennett's (born 1943) of a statistically optimal estimator for free-energy differences by leveraging overlapping ensembles and minimizing variance, now known as the Bennett Acceptance Ratio (BAR) method.[18] Unlike traditional FEP, which often suffered from poor sampling efficiency and variance amplification in low-overlap regimes, Bennett's method was derived from a variational principle that minimizes the statistical variance of the free-energy estimate by optimally reweighting configuration data from both ensembles. Instead of relying exclusively on either the reference or the target ensemble, the BAR method uses a balanced sampling strategy: Bennett showed that the optimal allocation of computational effort is to divide sampling time approximately equally between the two ensembles, which maximizes statistical efficiency under constrained computational budgets. These ideas were extended by Alan Ferrenberg and Robert Swendsen at Carnegie Mellon University to process the data from biased simulations.[58] This ultimately led to the so-called Weighted Histogram Analysis Method (WHAM),[59] introduced more than a decade after the introduction of BAR. Bennett's contribution marked a turning point in the development of free-energy estimation techniques; it established a rigorous mathematical framework that improved both accuracy and computational efficiency. His work reframed ensemble reweighting as a problem of statistical inference, thereby influencing both the practical execution of simulations and the broader theoretical discourse on the optimal allocation of computational resources in molecular modeling.

By the late 1970s, researchers had assembled a heterogeneous but increasingly coherent repertoire of tools for estimating free-energy differences in molecular systems. BAR addressed the problem of efficiency in comparing neighboring ensembles with sufficient phase-space overlap, while umbrella sampling enabled controlled exploration along predefined reaction coordinates to overcome rare-event barriers. Each method responded to the computational bottlenecks and representational constraints researchers faced at the time—whether the need to define discrete alchemical states or to traverse continuous configurational pathways. These methods reflected a deeper rethinking of how to structure simulations theoretically—how to intervene in sampling, where to allocate computational effort, and how to manage the trade-offs between bias and variance; in other words, how to organize molecular simulations with regards to the acquisition and evaluation of knowledge. This marked an inflection point in simulation practice: one where cunning statistical design and heuristic modulation began to replace raw sampling as the foundation for thermodynamic inference. These developments were soon going to be confronted to new challenges when extended to "realistic" models of molecular liquids and complex biomolecular systems.[60-63]

It was in this context that a study by Mihaly Mezei (born 1944) and David Beveridge (born 1938) at Hunter College of the City University of New York made a significant contribution by applying MC simulation techniques to compute the free energy of liquid water—an archetype of molecular complexity.[60] Their 1978 study represents one of the earliest attempts to use Kirkwood thermodynamic integration (TI) to evaluate Helmholtz free-energy differences for a dense, interacting molecular fluid. Using a high-fidelity ab initio potential based on configuration interaction data, they incrementally scaled the interaction potential via a coupling parameter λ, defined as:

$$U(\mathbf{r}_1, \ldots, \mathbf{r}_N; \lambda) = \lambda U(\mathbf{r}_1, \ldots, \mathbf{r}_N) \qquad \text{(Eq. 5)}$$

By integrating over λ from 0 to 1, they transitioned the system from a non-interacting reference (ideal gas at liquid density) to a fully interacting liquid phase. The resulting function $U(\lambda)$, computed at discrete intervals via canonical ensemble MC simulations on 64 water molecules with periodic boundary conditions, was numerically integrated using Gaussian quadrature. While the resulting free-energy estimates deviated from experimental values due to limitations in the underlying pairwise potential, the calculated entropy—obtained indirectly via thermodynamic relationships—showed notable agreement with empirical data. This partial success underscored both the promise and the constraints of applying Kirkwood TI to chemically realistic systems, especially when combined with simplified interaction models.

In another significant study, Johan Postma, Jan Haak, and Herman Berendsen (**Figure 3**) at the University of Groningen published in 1982 a foundational investigation to quantify the free-energy cost of creating repulsive cavities in water using MD simulations.[61] A cavity, in this context, denotes an excluded-volume region from which solvent molecules must be displaced to accommodate a solute–an energetically demanding process central to understanding solvation thermodynamics and hydrophobic hydration. Postma et al. employed Zwanzig's FEP method under isothermal–isobaric conditions to compute the Gibbs free-energy change associated with cavity formation. The interaction between water and a spherical, repulsive solute was modulated using a coupling parameter λ, varied from 0 (non-interacting) to 1 (fully interacting). The resulting free-energy difference $\Delta G$ was computed by integrating the ensemble-averaged derivative $\langle \partial U / \partial \lambda \rangle_\lambda$, following Eq. (2). This allowed for accurate estimates of free-energy changes for cavities with radii up to 0.32 nm, a scale at which hydrogen bonding and packing constraints are both relevant. An interesting aspect of the study by Postma and Berendsen was its direct comparison with scaled-particle theory (SPT), which models solvation using geometric approximations and statistical mechanics.[64] Viewed in historical terms, the study exemplifies how molecular simulations began to move beyond parameter testing toward model evaluation.

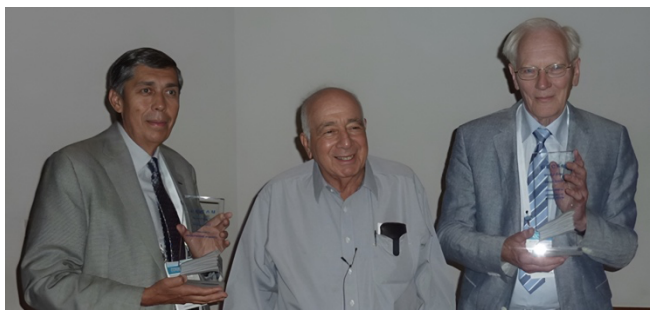

**Figure 3.** From left to right: Jean-Pierre Hansen (born 1942), Berni J. Alder (1925–2020), and Herman J. C. Berendsen (1934–2019), photographed during the 2013 CECAM Berni Alder Prize ceremony at the XXV IUPAP Conference on Computational Physics in Moscow. The award jointly recognized Hansen and Berendsen for foundational contributions to molecular simulation: Hansen for his contributions to liquid-state theory and coarse-graining approaches; Berendsen for pioneering methodologies in biomolecular dynamics, including isothermal molecular dynamics and constraint-based integration schemes that helped establish simulation as a key tool in structural biology and biophysics. The image documents an encounter between leading figures whose work shaped distinct domains of simulation science: Alder, a founder of molecular dynamics; Hansen, a central architect of statistical-mechanical theories of liquids; and Berendsen, a principal developer of simulation protocols for biological systems. Photograph courtesy of CECAM. Resolution improved via AI treatment.

Concurrently, Arieh Warshel (born 1940) formulated a theoretical approach exploiting the Zwanzig FEP method to determine the electrostatic effect of a fluctuating polar solvent on the activation free energy in electron- and proton transfer reactions occurring in solution.[62] In 1984, the FEP theory was expanded to describe electron- and proton transfer reactions occurring in enzymes.[63] Warshel's application of free-energy methods were primarily concerned with energetic effects associated with a redistribution of electrostatic charges, which dominate these types of reactions.[ix] Over the following years, this formulation would be pursued in additional studies of enzymatic reactions.[65,66]

These contributions marked a broader transition in the late 1970s and early 1980s, as simulations shifted from idealized fluids to chemically and biologically relevant systems. While Mezei and collaborators had shown that Kirkwood TI could be adapted to a high-density molecular liquid and Postma and colleagues applied this framework to the domain of nonpolar solvation, Warshel used Zwanzig's FEP to study electrostatic effects in electron- and proton transfer reactions in enzymes. In doing so, these studies embedded molecular simulations more firmly within the evolving methodological ecology of statistical mechanics and physical chemistry.

## 4. EMERGENCE OF ALCHEMICAL TI AND FEP IN COMPUTATIONAL BIOLOGY

Methods for calculating free energies in molecular simulations underwent a substantive transformation during the 1980s, as researchers sought to confront the challenges posed by increasingly complex biochemical systems. Researchers such as J. Andrew McCammon (born 1947) and William Jorgensen (born 1949) played a major role in translating theoretical frameworks into biologically meaningful applications.[21,22,67-71] In 1984, Tembe and McCammon[21] introduced the concept of alchemical FEP in which free-energy differences are estimated by statistically perturbing between chemically distinct molecular species. This formulation proved particularly effective in constructing FEP protocols for ligand–receptor binding interactions and solvation thermodynamics, as further refined in subsequent work.[67,68] Concurrently, Jorgensen and Ravimohan[22] applied FEP using discrete values of the coupling parameter λ and established a protocol in which the Hamiltonian is smoothly interpolated between solvation environments. Their study, focusing on the hydration free-energy difference between ethane and methanol, demonstrated the method's capacity to model transformations across chemically diverse solutes with predictive accuracy. These contributions redefined the operational scope of free-energy methods and expanded their application from idealized liquids to molecular transformations central to biochemistry and pharmaceutical design.

The idea of adapting of alchemical transformations to macromolecular systems reportedly occurred to J. Andrew McCammon, then at the University of Houston (**Figure 4**), during a scientific meeting,[x] where he was inspired by Berendsen's presentation on hydrophobic cavity simulations.[69] In their earlier work, Postma and colleagues had demonstrated the feasibility of using natural TI and FEP to compute free-energy differences for cavity formation in water. The success of these methods in simple systems informed McCammon's insight that they could be adapted to biologically relevant interactions such as enzyme-inhibitor binding. This

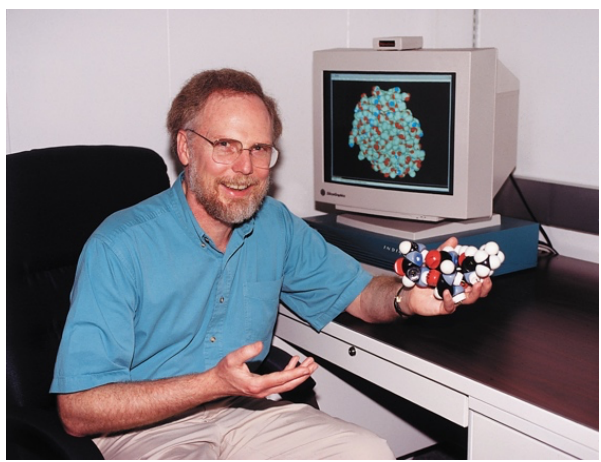

**Figure 4.** J. Andrew McCammon (born 1947), computational chemist and currently professor at the University of California, San Diego. Pioneered biomolecular molecular dynamics simulations and early applications of alchemical free-energy perturbation to protein–ligand binding. Photographed in 1994 at the University of Houston holding a model drug candidate for cancer research (courtesy of J. Andrew McCammon).

moment of recognition is vividly recalled by McCammon:[69]

> My focus on basic research expanded in the early 1980s when the wife of a close colleague developed cancer. I became very interested in the possible use of molecular dynamics simulations for drug discovery. How could one calculate relevant thermodynamic quantities, such as the free energy of binding an inhibitor to an enzyme? An answer came during another CECAM meeting in Orsay, in July 1983.[x] Herman Berendsen was describing the use of thermodynamic perturbation theory to calculate the change in the free energy of spherical cavities in water with increasing cavity radius. It was a warm afternoon; my mind was wandering a bit. However, suddenly I visualized the use of thermodynamic perturbation theory to calculate the change in free energy of an enzyme-inhibitor complex, when the inhibitor is changed into a slightly different molecule while bound to the enzyme. With a corresponding calculation for the inhibitors in water, one could use thermodynamic cycle arguments to compute the relative binding strengths of the inhibitors—something that should be useful in drug discovery. Bhalu Tembe, a postdoc, soon completed a demonstration simulation with an extremely simple model. It was rejected by *Nature*, for lack of general interest, and published in an obscure computational chemistry journal in 1984.[69]

This idea was soon operationalized in the 1984 study by Tembe and McCammon, where they applied FEP methods to the case of ligand-receptor binding.[xi] The one-sentence abstract of their article stated the objective clearly: "A simple theoretical approach is outlined for calculating differences in the free energy of binding of related ligand–receptor pairs."[21] While their contribution did not introduce new mathematical formalism and relied on 150 years old "technology" such as thermodynamic cycles, it vividly demonstrated how existing theoretical tools could be creatively reconfigured for biochemical contexts—linking perturbation theory with molecular recognition.

At the time, molecular dynamics simulations were increasingly applied to biochemical systems, but the computation of binding free energies essentially remained an inaccessible problem. McCammon's contribution was to envision that the machinery of statistical mechanical perturbation theory, already tested on solvation and cavity problems, could be extended to macromolecular binding via thermodynamic cycles. His proposal was not simply a formal innovation; it reimagined the structure of simulation workflows and introduced new possibilities for predicting biochemical affinity in silico.

In practice, the conceptual basis of thermodynamic cycles can be succinctly stated as follows: free energy is a state function and, as such, the total free-energy change $\Delta G$ around any closed, reversible cycle must be zero. This property enables the use of non-physical or alchemical transformations—such as gradually morphing a methyl group into a hydroxyl group, annihilating an atomic substituent, or interpolating Lennard-Jones parameters between distinct species—none of which are realizable in laboratory experiments (**Figure 5**). The relative binding free energy of two inhibitors, defined as the free-energy difference between their bound and unbound states, can thus be evaluated indirectly using a cycle of computational transformations.

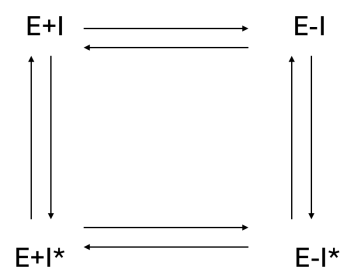

**Figure 5.** Thermodynamic cycle for calculating the relative binding free energy between two inhibitors, I and I*, to a common enzyme. The horizontal legs represent physical binding processes (E + I → E–I and E + I* → E–I*), while the vertical legs denote alchemical transformations between I and I*, carried out separately in the bound (top to bottom, left) and unbound (top to bottom, right) states. By thermodynamic consistency, the sum of free-energy changes around the closed cycle is zero, allowing the relative binding affinity to be computed from the difference between the two vertical (alchemical) free-energy changes: $\Delta G_{binding}(I^*) - \Delta G_{binding}(I) = \Delta G_{bound}(I \rightarrow I^*) - \Delta G_{unbound}(I \rightarrow I^*)$.

McCammon's specific interest was to determine the relative binding affinities of two inhibitors, I and I*, to a given enzyme. This situation involves four thermodynamic states: the enzyme bound to I (E–I), the enzyme bound to I* (E–I*), and the two corresponding unbound forms (E+I and E+I*), as depicted schematically in Fig. 5. By constructing a thermodynamic cycle that links these four states, the binding free-energy difference between I and I* can be expressed either as the difference in horizontal transitions (which are experimentally accessible binding events) or as the difference in vertical transitions (which correspond to alchemical transformations of I to I*). The algebra follows from the cyclic property: $\Delta G_{binding}(I) = G(E+I) - G(E-I)$, and $\Delta G_{binding}(I^*) = G(E+I^*) - G(E-I^*)$. Therefore, the difference in binding free energy is: $\Delta G_{binding}(I^*) - \Delta G_{binding}(I) = [G(E+I^*) - G(E+I)] - [G(E-I^*) - G(E-I)] = \Delta G_{bound}(I \rightarrow I^*) - \Delta G_{unbound}(I \rightarrow I^*)$. The latter correspond to the vertical transformations, which represent hypothetical, non-physical processes in which the molecular identity of the inhibitor is changed from I to I* in silico while maintaining the physical context (bound or unbound). By computing these alchemical free-energy changes using FEP or TI, one can estimate relative binding affinities without requiring direct simulation of the binding event itself. The significance of this formulation was twofold. First, it demonstrated that binding thermodynamics, long considered prohibitively difficult to simulate with atomistic detail, could be approached through formal manipulations of state functions. Second, it illustrated how computational models could leverage theoretical constructs such as thermodynamic cycles to produce observables with direct experimental counterparts. Although the vertical legs of the cycle do not represent physical processes, they offer a rigorous route for estimating free-energy differences between real biochemical states using transformations grounded in statistical mechanics.

Shortly after the publication by Tembe and McCammon, William Jorgensen and C. Ravimohan at Purdue University extended the methodological framework of alchemical FEP into the domain of solvation thermodynamics.[22] Their study computed the hydration free-energy difference between ethane ($CH_3$-$CH_3$) and methanol ($CH_3$-OH) using Metropolis MC simulations, representing one of the earliest implementations of alchemical perturbation for solute transformation in condensed-phase environments. The chemical change was mediated by a coupling parameter $\lambda$, which interpolated between the end states: at $\lambda = 0$ the solute was ethane, and at $\lambda = 1$ it was methanol. Rather than relying on sampling from a single reference state (as in Wong and McCammon's single-step formulation), Jorgensen and Ravimohan conducted multiple simulations at discrete $\lambda$ values, incrementally advancing the transformation. This piecewise procedure resembles the earlier approach used by Mezei et al. for liquid water, albeit using FEP instead of TI, essentially establishing the modern FEP protocol that is commonly used to this day.

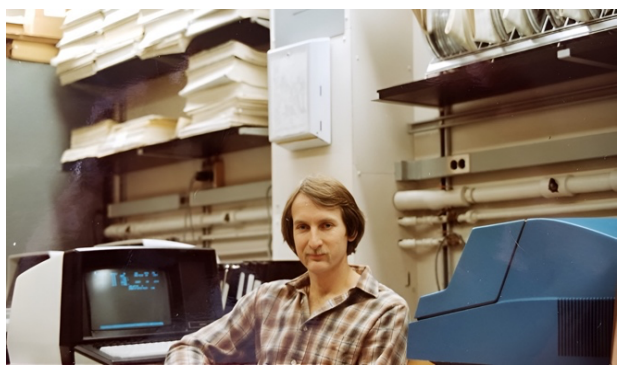

**Figure 6.** William L. Jorgensen (born 1949), computational chemist and currently professor at Yale University. Photographed in 1982 in his laboratory at Purdue University, seated at a Harris H80 minicomputer terminal, with shelves behind him filled with output binders and reels of magnetic tape—typical data storage media in the early era of computational chemistry. At the time, Jorgensen was finalizing the custom MC code for liquid-state simulations and alchemical free-energy calculations that he would later develop into the BOSS (Biochemical and Organic Simulation System) program (courtesy of William Jorgensen). Resolution improved via AI treatment.

A core methodological challenge in this case arose from the unequal number of particles in the two end states, an issue that Tembe and McCammon had acknowledged but deferred, noting that "the case in which the number of atoms is changed, will be presented elsewhere."[21] In contrast, the issue could not be avoided in the simulation study of Jorgensen and Ravimohan.[22] Adopting a united-atom representation, ethane consisted of two Lennard-Jones interaction sites (representing $CH_3$ groups), while methanol included an additional site for the hydroxyl hydrogen. To preserve consistency in the number of degrees of freedom across the simulation, the authors introduced a "dummy" or "ghost" non-interacting particle in the ethane state to represent the hydroxyl hydrogen. This particle, initially inert, was smoothly activated as $\lambda$ increased, gradually acquiring both partial charges and Lennard-Jones parameters until it became fully interacting at $\lambda = 1$. This strategy ensured that the alchemical transformation was mathematically well-defined and energetically stable throughout the trajectory.

The transformation was implemented using Jorgensen's custom MC simulation code (**Figures 6** and **7**), with configurational sampling governed by a Metropolis energy-based acceptance criteria.[xii] The dummy hydrogen atom, corresponding to the hydroxyl proton in methanol, was initially held fixed and non-interacting at $\lambda = 0$. As $\lambda$ increased, interaction parameters—van der Waals and electrostatic terms—were progressively introduced, enabling a smooth transition to the methanol state. Jorgensen remarked, "In Monte Carlo, the atoms move only if we allow them," a statement reflecting the method's interpretive control and the granularity of configurational oversight.[xiii] This staged activation facilitated the interpolation between chemically distinct solutes while preserving the ensemble compatibility and physical coherence of intermediate states. This wasn't just a code-level or implementation-level fix; it reframed how to preserve mathematical continuity in systems with topological disparity (e.g., different numbers of atoms). In embedding a reversible transformation within a single simulation framework, Jorgensen and Ravimohan demonstrated how perturbation theory could be extended, both mathematically and operationally, to transformations of real chemical interest.

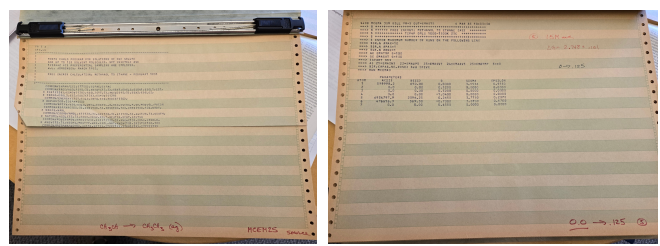

**Figure 7.** Original line-printer output of the FORTRAN Monte Carlo code used by William L. Jorgensen for the free-energy calculation of the methanol-ethane alchemical transformation in solution, as reported in the 1985 paper with C. Ravimohan. The code header identifies the simulation as a solute-in-solvent system in the NPT ensemble with preferential sampling for enhanced configurational exploration, while the handwritten annotation ($CH_3OH \rightarrow CH_3CH_3$) at the bottom marks the direction of the transformation. This routine, labeled "MCEM25", formed part of the early development of the BOSS (Biochemical and Organic Simulation System) program and illustrates the operational context in which early alchemical free-energy perturbation methods were implemented on punch-card systems in the 1980s (courtesy of William Jorgensen).

The subsequent two articles by Wong and McCammon[68] and by Lybrand, McCammon and Wipff[67] applied FEP directly to investigate ligand binding in biological targets. Whereas the initial publication by Tembe and McCammon[21] had illustrated the concept of using alchemical FEP to study ligand binding with simplified model systems, Wong and McCammon[68] computed the relative binding affinities of benzamidine and p-fluorobenzamidine to trypsin and compared the interaction of benzamidine with wild-type and mutant forms of trypsin—where a Glycine residue at position 216 was substituted with Alanine. This work enabled the demonstration of relative binding free-energy estimation between chemically similar inhibitors in an enzymatic context, a methodological milestone for rational drug design.

The alchemical FEP calculations were performed using a single MD trajectory of the wild-type enzyme-inhibitor complex, which served as the reference state. All alchemical substitutions were introduced by modifying the atomic identities within configurations sampled from this trajectory; this avoided the need for separate simulations of the mutant systems and reduced computational cost. The simulations employed the GROMOS package[74] and a united-atom force field in which aliphatic carbon-hydrogen groups are treated as single interaction centers, a common approximation in the 1980s that simplified the potential energy function and reduced the number of degrees of freedom.

In the case of enzyme mutation, the transformation from glycine to alanine at position 216 was carried out by inserting a $C_\beta$ atom directly into the configurations generated from the reference simulation. The relative free energy was then estimated using a single-step FEP expression:

$$e^{-\Delta G/k_B T} = \langle e^{-[U_1 - U_0]/k_B T} \rangle_0 \quad \text{(Eq.6)}$$

where the subscript 0 denotes averaging over the canonical ensemble generated with the original potential energy function $U_0$. This expression reflects the classic Zwanzig formula for exponential reweighting. In bypassing the construction of a $\lambda$-dependent interpolated Hamiltonian, Wong and McCammon avoided the numerical challenges associated with implementing staged alchemical pathways or extended thermodynamic integration. Instead, they relied on reweighting the ensemble generated under $U_0$, which assumes sufficient configurational overlap between the initial and final states. This methodological shortcut was feasible given the relatively minor perturbations involved in their transformations, i.e., small chemical differences between benzamidine and its p-fluoro analog or conservative amino acid mutations. While such approximations limit accuracy in more complex systems, the study convincingly demonstrated that relative binding affinities could be estimated using ensemble reweighting from a single trajectory. In a separate study,[67] Lybrand, McCammon and Wipff investigated the relative binding free energy of the anions $Cl^-$ and $Br^-$ to the macrotricyclic receptor SC24 in water. The calculation used the Zwanzig FEP formula with 10 intermediate values of $\lambda$, more or less in line with the simulation protocol introduced by Jorgensen and Ravimohan.[22] Alchemical FEP had previously been used to calculate the difference in solvation free energy for these anions.[75] Interestingly, the masses of the hydrogen atoms were artificially increased to 10 atomic mass units to allow the use of a larger dynamics timestep of 4fs, a strategy originated by Bennett[76] that is frequently used in present day calculations.[77] The two studies[67,68] showed that FEP could be a predictive tool in chemistry, biochemistry, and pharmacology. The groundwork had been laid down for a rigorous alchemical free-energy simulations strategy for drug design and ligand optimization.

## 5. IMPLEMENTATION OF FEP AND TI IN BIOMOLECULAR SIMULATION PROGRAMS

The pioneering studies by McCammon and Jorgensen between 1984 and 1986 generated significant momentum across the biomolecular simulation community and sparked a decisive shift toward applying free-energy methods to biologically realistic systems. In the following years, computational frameworks for TI and FEP were systematically incorporated into major biomolecular simulation platforms.

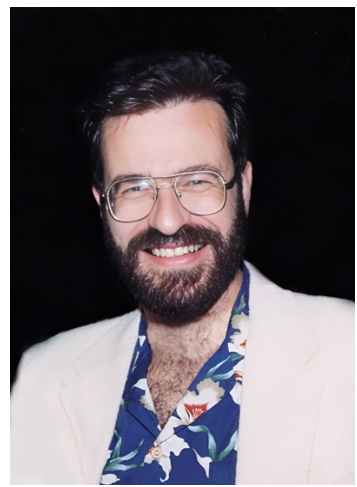

**Figure 8.** Peter A. Kollman (1944–2001), computational chemist and professor at the University of California, San Francisco. He contributed to the development of the AMBER MD package and to the application of alchemical free-energy perturbation methods for studying chemical and biological systems. Photograph taken in the early 1990s Photograph taken in the early 1990s (courtesy of Elsevier Science Ltd., from Fred E. Cohen, Structure 9, 885–886, 2001). Resolution improved via AI treatment.

An illustrative case is the AMBER (Assisted Model Building with Energy Refinement) program,[78] where early developments were catalyzed by private correspondence between William Jorgensen and his long-standing colleague and friend Peter Kollman (**Figure 8**), even before the publication of the Jorgensen-Ravimohan study.[22] Directly inspired by a pre-print of this manuscript (**Figure 9**),[22] Kollman's group at the University of California, San Francisco, initiated the implementation of alchemical methodologies into AMBER around 1984–1985.

According to Piotr Cieplak (born 1954), who was a visiting researcher in Kollman's group from 1985 to 1987 while he was a PhD student in the group of Prof. Wlodzimierz Kolos from the Faculty of Chemistry at Warsaw University, the earliest implementation of the free-energy perturbation into AMBER, including the coding of the early FEP modules was done by U. Chandra Sing, who then was a postdoctoral researcher in Kollman's group, [xiv] The implementation was based on technical strategies that essentially followed the single-topology transformation framework used in the Jorgensen-Ravimohan manuscript,[22] which had been communicated to Kollman by Jorgensen before its publication. According to David Pearlman, a later contributor to AMBER development, major coding milestones for free-energy capabilities had already been substantially completed by mid-1985. [xv] A personal letter from Kollman to Jorgensen dated June 27, 1985—only weeks after the submission (May 21, 1985) and acceptance (June 12, 1985) of the Jorgensen-Ravimohan manuscript—confirms that functional alchemical capabilities were operational within AMBER at that time (**Figure 9**). Interestingly, there is also a letter dated from 3 May 1985 from Martin Karplus, acknowledging this work prior to its publication (**Figure 9**).[xvi] Jorgensen knew McCammon and Karplus from his PhD at Harvard in the 1970's,[xvii] and was a very close personal friend of Kollman (**Figure 10**). These letters are revelatory of the close communication that was often present in the field at the time.

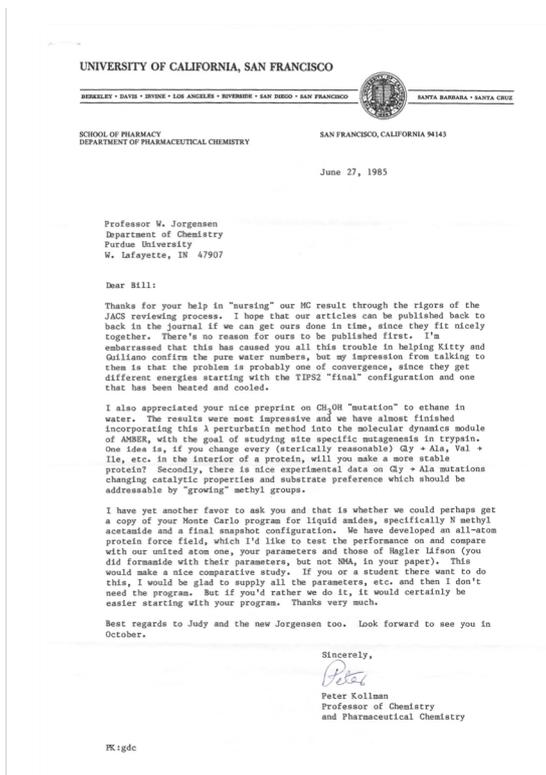
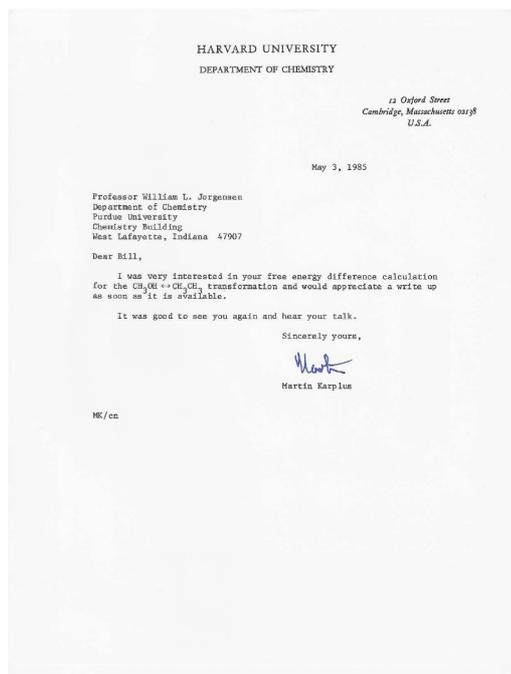

**Figure 9.** Letters from Peter A. Kollman and Martin Karplus (1930–2024) to William L. Jorgensen concerning the ethane-methanol hydration free-energy difference calculation later published by Jorgensen and Ravimohan.[22] (left) Letter from Peter Kollman, dated June 27, 1985 (University of California, San Francisco), reporting that λ-dependent perturbation methods were being integrated into the AMBER molecular-dynamics module and outlining planned applications to site-directed mutagenesis in trypsin. (right) Letter from Martin Karplus, dated May 3, 1985 (Harvard University), referring to the new work and requesting a pre-print when available. The two letters record correspondence on alchemical free-energy methods in 1985 (courtesy of William L Jorgensen).

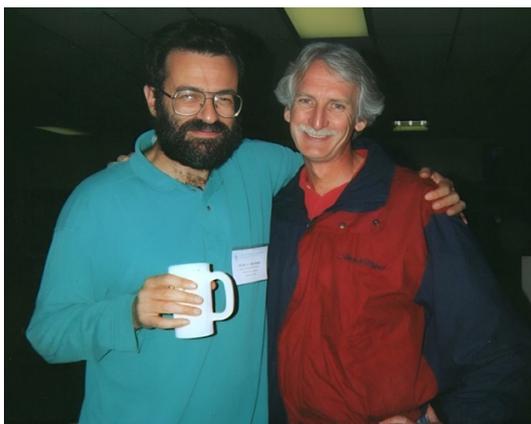

**Figure 10.** Peter A. Kollman (1944–2001) and William L. Jorgensen (born 1949) at the Computational Chemistry Gordon Research Conference (GRC) in 1996. The meeting brought together leading figures in theoretical and computational chemistry, including key contributors to the development and application of alchemical free-energy methods. This photograph documents the close friendship between two principal developers whose work shaped both the methodological foundations and practical implementations of free-energy methodologies in biomolecular simulations (courtesy of William L. Jorgensen). Resolution improved via AI treatment.

Consistent with AMBER's broader orientation toward the simulation of proteins, nucleic acids, and their interactions, the design of free-energy routines emphasized generality, modularity, and compatibility with force-field conventions. A pivotal conceptual decision was the adoption of a "single topology" framework: all atoms present in either molecular state were represented continuously in a shared coordinate set, with only non-bonded terms (Lennard-Jones potentials and Coulombic interactions) being subjected to λ-dependent scaling. Bonded terms (bond lengths, angles, and torsions) were preserved across λ, including for dummy particles introduced during transformations. This strategy privileged chemical continuity and configurational stability over a strict mapping of isolated physical states. It repositioned simulation as a constructive modeling practice, not merely a numerical reproduction of experimental endpoints. In maintaining internal connectivity, the approach minimized distortions in molecular geometry, reduced sampling artifacts near endpoints, ensured smoother energy landscapes across λ-space, and enhanced the numerical stability and convergence of free-energy estimates. Piotr Cieplak, recalls how the single topology method was implemented at the time,[xiv]

> In AMBER, the perturbation of one molecule into another one was performed in such a way that one needed to build a single topology for a hybrid molecule containing all necessary "ghost" atoms, which would grow to real atoms while other real atoms would disappear by transforming them into new ghost atoms. Alternatively, if the above was not possible or convenient, one needed to use thermodynamics cycle to first disappear one molecule and then grow a new molecule afterwards in its place. As I understand, this later approach did not work well at the time, due to instabilities occurring when large molecules needed to disappear or grow in solution or in a binding site. However, this method works well in the current AMBER program implementation.[xiv]

The adoption of the single-topology framework in AMBER expanded the scope of alchemical transformations, allowing molecular systems to undergo controlled changes without introducing discontinuities in their structural representation. Through stabilization of solute modifications and preservation of solvent integrity during transitions, the method supported robust free-energy calculations in chemically diverse scenarios. This framework provided the foundation for a series of influential studies by Kollman and collaborators,[79-81] which helped establish alchemical simulation as a predictive tool in biomolecular modeling.

Of particular note, Rao et al.[81] applied FEP to predict both the change in binding free energy and the catalytic activation barrier for a subtilisin mutant and achieved striking agreement with later experimental measurements. This study provided one of the earliest demonstrations that alchemical simulation could reproduce and anticipate experimental observables, which reinforced simulation as a predictive theoretical and computational tool rather than a retrospective fitting device. The implementation of FEP and TI within AMBER marked a transition in the epistemology of molecular simulations in the sense that the MD trajectory is not merely seen as a (classical) mimicry of reality, but as a sophisticated computational tool able to deliver rigorous free-energy estimates relevant to the real world through the simulation of some fairly abstract non-physical intermediates. This transition exemplifies how computational practice evolved through a dual negotiation of theoretical formalisms and pragmatic coding architectures, with molecular simulation emerging as an autonomous mode of inquiry at the crossroads of chemistry, physics, biology, and computation.

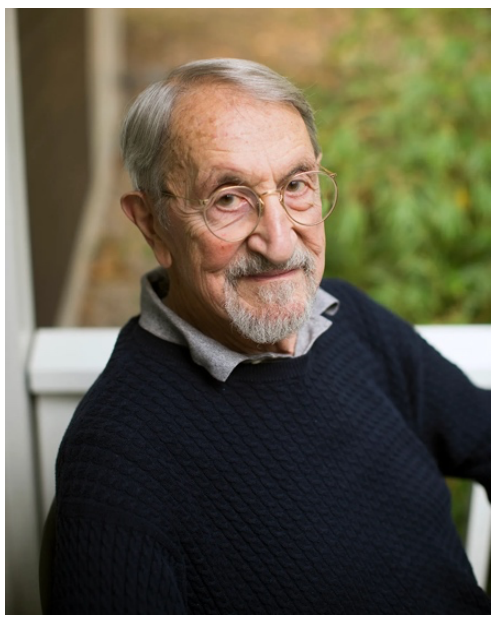

**Figure 11.** Martin Karplus (1930–2024), theoretical chemist and professor at Harvard University. Awarded the 2013 Nobel Prize in Chemistry (with Michael Levitt and Arieh Warshel) for the development of multiscale models for complex chemical systems. He contributed extensively to statistical mechanics and quantum chemistry, with a particular focus on molecular dynamics simulations, including applications of free-energy methods. Photograph taken around 2014 (courtesy of the Harvard Gazette and Marci Karplus).

Almost at the same time (1985–1987), Martin Karplus (**Figure 11**) and his team at Harvard University embarked on the implementation of alchemical TI and FEP methods into the MD program CHARMM (Chemistry at HARvard Macromolecular Mechanics).[xviii] On an interesting note, Karplus had had a direct and personal history with John Kirkwood, having almost done his PhD thesis with him when he was a graduate student at Caltech.[xix] With the arrival of postdoctoral researchers Jiali Gao (born 1962) from Jorgensen's group and Paul A. Bash from Kollman's group, lively discussions on free-energy calculations abounded from all corners within the Karplus group, creating a fertile intellectual environment for rethinking how thermodynamic cycles and perturbative methods could be operationalized within MD simulations.

The initial coding of what became the first widely used general-purpose free-energy module in CHARMM, later known as the BLOCK module, was primarily undertaken by Bruce Tidor (born 1961), a Ph.D. student under Karplus. Tidor, who had been an undergraduate at Harvard and had pursued further studies at Oxford, joined the Karplus group in the fall of 1985. Drawing on both prior discussions and emerging needs in molecular modeling, Tidor's work focused on creating a flexible, extensible codebase for diverse alchemical transformations within the framework of the CHARMM program.[82-84]

In contrast to AMBER's adoption of a single-topology model, the CHARMM BLOCK module implemented a dual topology approach. In this design, the atoms representing both the initial and final states coexisted explicitly within the simulation system and were coupled independently to the environment. This separation allowed greater flexibility when dealing with transformations involving structurally or electronically distinct end states. According to Gao,[xx] an additional motivation for the dual-topology architecture may have been to build a free-energy framework adaptable to hybrid QM/MM simulations, where dual-topology descriptions of the end states are inherently required.[86,87]

The initial version of the BLOCK module did not allow for granular control of molecular transformations with respect to individual energy terms. All contributions (bonded and non-bonded interactions) were scaled simultaneously over the course of a FEP or TI calculation for processes such as amino acid mutations. As the BLOCK facility was progressively enhanced over the years, it became possible to modulate both bonded and non-bonded interactions individually modulated as a function of the coupling parameter λ. However, the inclusion of bonded terms presented substantial methodological challenges. Unlike non-bonded interactions, which primarily modulate long-range energy contributions, bonded parameters govern the local structural framework of molecules. Ensuring the stability and physical consistency of intermediate states therefore required careful parameter tuning and rigorous validation, particularly near λ endpoints.

The full decoupling of dummy particles from both bonded and non-bonded interactions in the CHARMM free-energy implementation was utilized in the pioneering protein stability calculations by Gao and colleagues.[39] Using a dual topology framework, they rendered dummy particles functionally analogous to an ideal, non-interacting particle ensemble, preventing any unintended perturbations to the simulated system. The rationale for this design was to eliminate biases in the calculated free energies that could arise even from minimal residual interactions involving dummy particles. In their thermodynamic analysis of the hemoglobin interface mutant AspG1(99)β → Ala, the authors uncovered that individual residue and solvent contributions to the free-energy change could be extremely large, sometimes exceeding 60 kcal/mol, despite the net free-energy change between the normal and mutant hemoglobins being relatively modest (-5.5 kcal/mol, compared to an experimental value of -3.4 kcal/mol).[39] This result revealed the intricate cancellation effects underlying mutational free-energy changes in biological macromolecules.

Around the same period, the program GROMOS had not yet incorporated free-energy methods.[xxi] Early TI and FEP calculations in Groningen were performed using ARGOS,[xxii] a simulation program specifically developed by Tjerk P. Straatsma to address computational challenges associated with free-energy estimation.[88] The implementation of alchemical perturbation techniques in ARGOS was initiated around 1984, culminating in the 1986 publication by Straatsma, Berendsen, and Postma.[90] ARGOS supported a wide range of alchemical transformations using both perturbation and thermodynamic integration schemes. Meanwhile, Wilfred van Gunsteren and co-workers initiated systematic investigations into free-energy decomposition and its conceptual limitations, recognizing that individual energetic components do not constitute true state functions.[91,92]

The expansion of free-energy methods in the mid-1980s generated intense interest across the biomolecular simulation community.[xxiii] Although expectations were high, many discussions revealed persistent difficulties: free-energy calculations frequently failed to converge, decomposition strategies raised conceptual ambiguities,[91,92] and the treatment of dummy particles remained unsettled. These meetings captured a transitional moment in simulation practice, when optimism about methodological innovation was tempered by a growing awareness of technical and theoretical limitations.

## 6. CONSOLIDATION AND STABILIZATION (FROM THE LATE 1980'S TO 2000)

### a) Handling the Concept of Dummy Particles

Despite the foundational contributions of McCammon and Jorgensen, the apparent simplicity of their early model systems masked a deeper structural difficulty inherent in alchemical transformations: the management of dummy particles. These inert or decoupled atoms, introduced to interpolate between chemically distinct species, posed little practical difficulty in simple cases such as ethane-to-methanol transformations using united-atom models. Non-bonded interactions could be switched off cleanly, and free-energy differences computed in a straightforward single-step calculation based on Eq. 3.

In specific studies, such as perturbing the binding affinity of a monoatomic ligand in host-guest systems,[67] the issue of dummy atoms could simply be avoided. However, as free-energy calculations were extended to more complex biomolecular systems, however, the lack of a systematic strategy for handling decoupled atoms became increasingly problematic (**Figure 12**). In simple systems, the entropy introduced by unrestrained dummy particles could be ignored; in biomolecular contexts, it became destabilizing. Without bonded constraints, these atoms diffused freely throughout the simulation volume, inflating the number of accessible microstates and artificially biasing computed free energies through an entropy contribution that grew logarithmically with system size, proportional to log(V). This uncontrolled entropic inflation exposed a critical fragility in early alchemical frameworks, especially as simulation targets grew more heterogeneous and chemically complex.

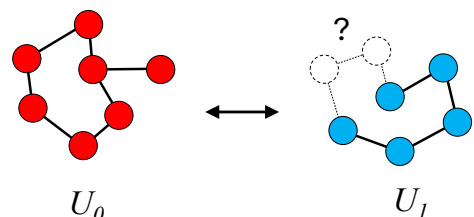

**Figure 12.** Schematic representation of alchemical free-energy perturbation between two states of potential energy $U_0$ and $U_1$ when number of interacting particles (red or blue) is different. When the physical particle counts are unequal, non-interacting "dummy" (or "ghost") particles—shown with dashed outlines—are added so that the coordinates are defined for both states. This ensures that the configurational integrals are evaluated in spaces of identical dimensionality. Correct management of these dummy particles and their connectivity is essential to avoid artifacts in the computed free-energy differences.

During the initial adaptation of FEP and TI methods into software packages like AMBER and CHARMM, early solutions emerged but remained partial and heterogeneous. Some implementations tethered dummy particles to their original positions with soft harmonic restraints, while others attempted to cancel internal degrees of freedom through selective switching of bonded and non-bonded terms. These pragmatic workarounds, while effective in limited cases, lacked formal generality and left simulations vulnerable to systematic artifacts. Inadequate constraint schemes or inconsistent interaction switching could easily introduce irreducible biases, undermining the thermodynamic consistency required for robust free-energy calculations.

By the late 1980s, it became increasingly recognized within the computational chemistry community that the naive dummy-particle treatment posed a fundamental obstacle to the reliability of alchemical methods. A decisive conceptual response emerged in the CHARMM implementation by Gao and colleagues.[39] In this framework, dummy atoms were stripped of both bonded and non-bonded interactions with the surrounding system, they exerted no forces and accumulated no energy. Consciously or not, uncoupled dummy atoms were modeled as ideal gas entities. This design minimized non-physical artifacts and sought to eliminate spurious contributions to the calculated free energy. However, it reflected a deliberate, new theoretical stance: dummy particles were no longer regarded as partial molecular fragments awaiting reconnection but as purely formal constructs—mathematical devices to facilitate smooth interpolation without carrying physical meaning. This reframing of dummy particles marked a critical step in the broader stabilization of alchemical free-energy methods. Treating non-physical intermediates as controllable formal artifacts, rather than as imperfect proxies for real molecular behavior, created a conceptual and technical foundation for later refinements, including dual topology models and rigorous treatment of dummy degrees of freedom within free-energy cycles.

The primary requirement for a valid alchemical free-energy calculation is that the overall free-energy difference around a closed thermodynamic cycle must accurately reflect the physical processes of interest. Achieving this consistency demands that all superfluous internal energy terms associated with dummy particles be fully switched off at both endpoints. In principle, such ideal cancellation ensures that dummy particles do not perturb the calculated free energies. Yet this idealization introduces a major difficulty: in the absence of bonded constraints, dummy particles behave as non-interacting particles and diffuse throughout the entire simulation volume V. An alternative strategy retains dummy particles as part of the connected topology, provided their energetic contributions cancel appropriately in the overall thermodynamic cycle. However, mishandling dummy particles (whether through inadequate restraints or inconsistent treatment of internal interactions) can introduce significant artifacts that compromise the physical accuracy and interpretability of computed free-energy differences.

These issues were well recognized by the developers of molecular dynamics programs. In the CHARMM implementation, exemplified by Gao and coworkers,[39] dummy particles were stripped of both bonded and non-bonded interactions with the rest of the system. The objective was to eliminate non-physical forces and prevent spurious energy contributions from distorting the calculated free energies. While elegant in principle, this approach reflected a specific methodological commitment: dummy particles were modeled as ideal gas entities, deprived of any physical function beyond their formal role in enabling the alchemical transformation.

Sampling non-interacting particles within this dual-topology paradigm, however, introduced practical difficulties, as Gao and colleagues themselves observed. In contrast, the AMBER development team adopted a different strategy. In personal communication with the authors, David Case—a long-time collaborator of Peter Kollman—explained that the use of a single-topology framework in AMBER was primarily motivated by a concern to avoid the entropic complications associated with ideal gas behavior at the endpoints.[xxiv] Within the single-topology framework, Pearlman and Kollman formulated a pseudopotential of mean force correction to account for bond stretching contributions.[93] David Pearlman confirmed that it was well understood within the group that internal interactions involving dummy particles in single-topology could perturb results and shift the thermodynamic endpoints.[xxv] Nevertheless, it was generally assumed that such contributions would be small and would largely cancel in practical calculations. AMBER therefore retained bonded energy terms for dummy particles to prevent their dispersal across the simulation box, even though this approach introduced approximations of its own.[xxv]

These different strategies across AMBER, CHARMM, and subsequent implementations had lasting consequences for the accuracy and scope of these computational approaches, and on the development of other simulation programs. Alchemical free-energy methodologies were implemented in GROMOS with a single-topology approach similar to AMBER.[xxi] NAMD adopted a dual-topology framework.[94] Over time, a dual-topology framework was also incorporated in AMBER, in addition to the single-topology framework.[95] On the other hand, CHARMM remained exclusively based on a dual-topology framework.[83] However, such technical decisions did not emerge in isolation. The consolidation of alchemical free-energy methods depended less on unified theoretical breakthroughs than on the interplay between infrastructural resources, local software practices, and institutional support structures. Access to high-performance computing resources, whether through national centers, departmental clusters, government-sponsored initiatives, or private collaborations, often determined whether a research group could pursue simulations at biologically relevant scales. Local forms of software expertise were equally decisive: in many laboratories, a single postdoctoral researcher or graduate student often acted as both scientific translator and software architect. The development of the field was uneven, shaped by conditions that extended beyond intellectual ambition or institutional prestige. Funding access, frequently aligned with national or industrial priorities (for example, U.S. NIH initiatives in structure-based drug design, DARPA programs for molecular simulation infrastructure, or pharmaceutical collaborations involving groups such as DuPont or Merck), influenced which types of simulations were pursued. Equally important were the pathways through which software was disseminated, modified, and adapted across different laboratories. Much of the operational knowledge that made free-energy simulations viable was tacit: embedded in handwritten notes, custom scripts, undocumented code modifications, and transmitted through face-to-face mentorship, personal correspondence, or direct apprenticeship. In many cases, the success of a free-energy calculation depended less on the availability of a published method than on the presence of someone capable of translating theoretical formulations into executable, machine-specific code.

Although earlier codes employed various ad hoc workarounds, a fully systematic framework for dummy particle treatment remained elusive throughout the 1980s and 1990s. A major conceptual advance came during the doctoral research of Stefan Boresch (born 1965) under Martin Karplus between 1990 and 1997.[96] Their analysis proposed the use of harmonic potentials to hold dummy particles near specified coordinates, which allowed their contribution to the partition function to be analytically isolated.[97-99] Boresch demonstrated that such restraints did not distort the computed free energies and offered a solution that combined formal theoretical soundness with practical computational strategies. As he later recalled, the motivation for this work arose directly from grappling with the unresolved ambiguities of earlier free-energy implementations and free-energy component analysis raised by van Gunsteren and co-workers.[xxvi] Boresch explains the context in which his work was developed,[xxvii]

> I was a PhD student in chemistry, and I joined the Karplus group in early 1991. My first project was an in-depth study on model systems (rather than proteins) to understand the validity of the free-energy component analysis and the source of unexpectedly large "self-term" contributions.[100] The major opposition to free-energy component analysis was Wilfred van Gunsteren expressed in papers co-authored with Alan Mark.[91] Pursuing this work led to the reevaluation of what free-energy components could and could not do.[102] More importantly, however, I realized that no one understood how changes in bonded energy terms contributed to single and double free energy differences. Others had presented results in this direction,[103,104] but not taking into account the distinction between a single and dual topology setup,[105] the conclusions of these studies seemed contradictory. When I discovered an old paper by Herschbach, Johnston and Rapp (HJR)[106] I realized that I had found the key. Together with a detailed treatment of coordinate transformations of the partition function in the collected works of Gibbs (which I read in an edited version by Donnan and Haas),[107] I could reproduce HJR's examples, as well as the data reported by Hermans,[104] first with paper and pencil, later by computer simulations. Two chapters of my thesis[96] are concerned with the role of bonded energy terms in free-energy simulations, depending on the use of single vs. dual topology; these resulted in three publications.[97-99] I defended my thesis on Nov 28, 1996 in Straßburg, so my official degree is from march 1997, but the main manuscript of my thesis were published only in 1999.[xxvii]

Boresch and Karplus offered one of the earliest formal demonstrations that, if the contribution of dummy particles without non-bonded interactions could be integrated and factored out of the configurational partition function, they would add only a constant term to the legs of a thermodynamic cycle, which cancels in double free-energy differences. S. Shobana (born 1964), Benoît Roux (born 1958), and Olaf Andersen (born 1945)[108] later showed that this reasoning remained valid as long as each dummy particle was connected by at most one bond, one angle, and one dihedral to the physical core. However, even after these advances, a fully general treatment of dummy particles in single-topology schemes with arbitrary connectedness would not be finalized for several years.[109]

*b) Absolute Binding Free Energy of Ligands*

The methods, exemplified by the works of McCammon and co-workers[21,68] and Jorgensen,[22] allowed researchers to estimate relative differences in solvation free energies and ligand–receptor binding affinities. Nevertheless, as molecular simulations became increasingly sophisticated, the need to calculate the so-called standard or "absolute" binding free energy of ligands (as opposed to a relative change in the binding free energy) gained scientific urgency. Absolute Binding Free-Energy (ABFE) methods were introduced to quantify the binding process comprehensively, and to extend thermodynamic analysis beyond purely comparative frameworks. This development was driven by specific challenges inherent in applying alchemical methods to model the energetics of complex biological systems, coupled with the goal of improving alignment between theoretical predictions and experimental observations.

The adoption of ABFE calculations coincided with a broader reorganization of methodological priorities in molecular simulation. Earlier practices emphasized formal derivations and analytical reproducibility, whereas by the early 2000s, researchers increasingly valued predictive reliability within experimental contexts such as Computer-Aided Drug Design (CADD), protein engineering, and materials science. Evaluation criteria moved toward performance in real-world applications; methods were assessed based on their capacity to produce consistent, actionable results under conditions of empirical uncertainty. Reliability came to be associated less with strict derivation from first principles and more with demonstrated operational stability across a range of modeling contexts. In this manner, the expectations associated with simulation practice were realigned to accommodate the complexities of biological function and therapeutic development. This section outlines the conceptual and technical adjustments that supported the broader application of ABFE methods between the late 1980s and early 2000s, including the resolution of challenges such as the design of restraint potentials and the incorporation of standard-state corrections.

Absolute binding free energy quantifies the total thermodynamic cost associated with the formation of a ligand-receptor complex from fully solvated, unbound states. In contrast to relative binding free-energy calculations, which measure differences between alternative ligands, ABFE reconstructs a full thermodynamic cycle. To offer a complete characterization of the binding event, the calculation is typically conceptualized as a thermodynamic alchemical process in which the ligand is progressively decoupled (or "annihilated") in the binding site and then rematerialized in the solvent. By formally constructing a thermodynamically closed cycle, the alchemical process seeks to yield an estimate of the standard binding free energy under defined conditions. The binding of a ligand L to a receptor R can be described by the equilibrium reaction:

$$R + L \xrightleftharpoons{K_{eq}} RL \qquad \text{(Eq. 11)}$$

where $K_{eq}$ is the equilibrium constant defined by the concentrations of free receptor [R], free ligand [L], and the complex [RL] at equilibrium. The equilibrium constant $K_{eq}$ is related to the standard free energy of binding, $\Delta G_{eq}^{(\circ)}$, by the expression:

$$\Delta G_{eq}^{(\circ)} = -k_B T \ln[K_{eq} C^{(\circ)}] \qquad \text{(Eq. 12)}$$

where $C^{(\circ)}$ denotes the standard concentration (typically 1 M). ABFE calculations held historical significance due to its capacity to offer a more complete energetic analysis; yet it posed substantial conceptual and computational challenges, particularly within the constraints of the computational infrastructures and sampling methodologies available during the late 1980s.

An early application of ABFE calculation was undertaken by Hermans and Shankar[110] at the University of North Carolina at Chapel Hill's Department of Biochemistry. Their study addressed the challenges associated with accurately reconstructing the complete thermodynamic cycle of ligand binding, focusing on the interaction of xenon with myoglobin. Myoglobin and xenon presented distinct challenges due to the subtle molecular rearrangements necessary for binding, coupled with the availability of crystallographic data[110] that provided a reliable structural framework. These conditions made the system a representative model for developing methods capable of describing the full thermodynamic cycle.

Hermans and Shankar developed an approach based on an early variant of alchemical free-energy perturbation (FEP). Their work highlighted the importance of achieving near-reversible free-energy pathways. To stabilize the non-interacting xenon atom and prevent its unphysical drift, they introduced a non-physical harmonic restraint that maintained the xenon close to its binding site within the protein structure.

As the perturbation parameter λ increased stepwise from 0 to 1, the harmonic restraint was progressively released, ensuring the xenon atom remained localized during the non-interacting phase. Importantly, Hermans and Shankar emphasized that "The harmonic restraint is in addition essential to give a well-defined reference state for the xenon atom when it does not interact with the protein, i.e., when λ=0."[110] At this endpoint, the xenon atom was effectively confined to a small volume, characterized by the integral of the Boltzmann factor of the harmonic potential. This strategy demonstrated their recognition that the restraint was essential for defining a meaningful reference state for a non-interacting gas-phase atom. Although Hermans and Shankar calculated the pressure corresponding to the ideal gas confined within this small volume, they did not extend their framework to define the standard state concentration necessary for ligands in solution, thereby limiting the broader applicability of their method. Moreover, because their reference state corresponded to a xenon atom in an ideal gas, their approach did not address solvation free-energy contributions critical for ligand binding in aqueous environments. It is interesting to note that, while Jan Hermans was primarily an experimentalist, he contributed significantly to the development of biomolecular simulations, notably with the theoretical formulation of absolute binding free energies[110,111] and the development of the widely used SPC water model in collaboration with Herman Berendsen (**Figure 13**).[112]

Further developments in absolute binding free-energy calculations emerged with the investigations by Cieplak and Kollman,[113] and independently by Jorgensen and co-workers.[70] These studies extended the framework to solvated ligands, thereby incorporating the energetic cost of desolvation into the thermodynamic cycle. Around the same period, McCammon and Rebecca Wade examined the hydration of specific cavities within proteins.[114] Although their focus was on water molecules rather than traditional ligands, the methodological structure was consistent with the principles underlying ABFE calculations, as the water molecules occupied discrete binding sites within the protein matrix.[xxviii]

In these early studies, the gradual removal of water molecules or ligands was implemented using a double annihilation strategy.[70] This approach revealed practical limitations, particularly in cases where the non-interacting species could drift away from its binding site during decoupling. Such behavior introduced irreversibility into the free-energy pathways: calculations carried out with the coupling parameter λ increasing from 0 to 1 produced results that differed from those obtained by decreasing λ from 1 to 0. These discrepancies highlighted the inherent instability of simulations where unrestrained decoupled species could diffuse freely. By the mid-1990s, concerns regarding the formal structure of double annihilation methods became increasingly prominent. Researchers recognized that the conceptualization of binding free energy as the difference between the free energy of a ligand in the binding site and that of the ligand in solution. While intuitively appealing, this representation of the problem obscured subtle complexities about the reference state that demanded a more systematic theoretical treatment. A formulation of the binding constant based on the PMF between the ligand and its receptor would have shed light on these issues.[xxix]

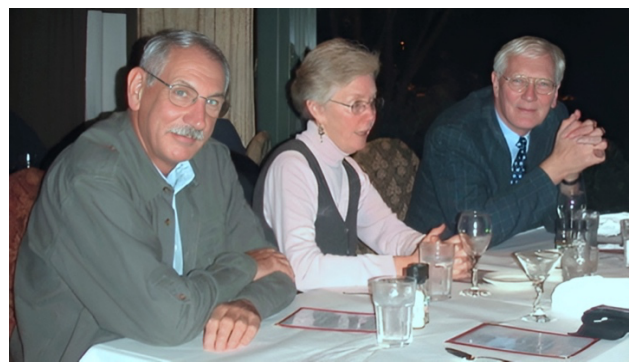

**Figure 13.** From left to right: Jan Hermans (1933–2018), Cynthia Hermans (his spouse), and Herman J. C. Berendsen at a conference in late 2003. Jan Hermans was a biophysicist at the University of North Carolina at Chapel Hill, known for contributions to protein structure and dynamics, including interpretations of electron density maps and the theoretical foundations of macromolecular simulations. The photo captures the collegial network among researchers who shaped early simulation methodologies (courtesy of Edward T. Samulski). Resolution improved via AI treatment.

Conceptual refinements in the treatment of binding free energy emerged from the work of Roux and collaborators,[116] who developed a formalism to quantify the occupancy of water molecules in the binding sites of bacteriorhodopsin (bR), a light-driven proton pump located in the purple membrane of Halobacterium halobium. Their analysis addressed the complexities associated with defining the thermodynamic reference state in the context of cavity hydration. One of the co-authors, Benoît Roux, later described the scientific setting in which this work was conducted:

> Mafalda Nina was a PhD student jointly mentored by Jeremy Smith who was at the CEA in Saclay, and myself at the Université de Montréal. Her thesis (1991-1994) was to study the structural and function of bacteriorhodopsin, a light-driven proton pump. Several lines of evidence indicated that water molecules played an important functional role. Every summer, Mafada would come to Montreal for a period of 3-4 months to work. I recall a long discussion in July 1994 with Mafalda and Regis Pomes, who was a postdoc in the lab, where were considering the water molecules in specific cavities in the neighborhood of the Schiff base. We wanted to go beyond simple statements about interactions energy and cavity size and were trying to find a way to say something quantitative about the probability of occupancy. We started to write equations on the board about the probability of occupancy, expressed as averages with configurational integrals involving an indicator function. After some relatively simple manipulations, in which we introduced an intermediate stage with a harmonic restraint on the water molecule ligand, we recognized (to our surprise and embarrassment) that the probability of occupancy could be expressed in the familiar form of first order binding equilibrium. While we had not initially started thinking along those lines, our analysis had led us to a formal statistical mechanical treatment for the equilibrium binding free energy. The FEP binding calculations were carried by Mafalda and Regis using the PERT module of CHARMM, and Mafalda presented the results at her thesis defense in December 1994 in Paris. After a long delay, the manuscript was finally published in August 1996. In retrospect, I believe that part of our good luck when developing our statistical mechanical formulation was that our formal starting point had been the probability of occupancy of the ligand water molecule rather than some traditional concept of binding free energy.

The methodological approach formulated by Roux et al.,[116] based on the application of a non-physical harmonic restraint to the decoupled ligand, provided a consistent strategy for defining the reference state, thereby supporting reversible free-energy calculations. Around the same time, Hermans and Wang[111] extended similar considerations to systems where ligands were in equilibrium with proteins in solution, again employing harmonic restraints to stabilize the ligand during decoupling and to define the thermodynamic reference state, and Gilson and collaborators[117] presented a formal statistical mechanical expression for the ABFE in complex systems with a correct treatment of the standard concentration, and used it to formulate the double decoupling method (DDM).[117] Gilson recalls the context in which this work was developed:[xxx]

> I was a postdoc with Andy McCammon at University of Houston from 1991 to 1993. One day at lab meeting, he stood up and told us he had just gotten back from consulting at Bristol-Myers Squibb, where he had been asked lots of questions about changes in rotational and translation entropy in protein-ligand binding, something called the cratic entropy, and related ideas. He held up a pile of papers and asked if anyone would be interested in working on these issues, and I said I'd be very interested. After a while, I noticed that there didn't seem to be any book or article that had $\Delta G_{eq}^{(\circ)}$ of binding on the left of the equals sign and an expression in terms of configuration integrals or partition functions on the right. I ended up taking this problem to my first faculty position at CARB in Rockville, Maryland, and continued to work on it with my first postdoc, James Given. James was strong in statistical thermodynamics and convinced me that it was okay to use an indicator function to define the bound complex. It was Bruce Bush, from Merck New Jersey, who, among other things, pointed out at a Gordon conference that absolute binding free energies are more properly standard binding free energies, which depend on the choice of standard concentrations, and that this dependency was missing from the absolute binding free-energy calculations being done at that time.[xxx]

The work of Roux et al.,[116] Hermans and Wang,[111] and Gilson et al.[117] collectively established a systematic foundation for the calculation of ABFE with a correct treatment of the standard concentration in solvated systems. Each contribution clarified the necessity of restraint-based reference state definitions and illuminated the implications of standard concentration dependencies. Subsequent studies by Boresch et al.[118] and Roux[119,120] expanded and generalized these principles, ensuring that binding free-energy calculations could be carried out in a reversible and thermodynamically consistent manner. By the early 2000s, standardized computational protocols for ABFE methods had been consolidated and documented,[121] thereby facilitating their broader application in computational chemistry and related disciplines.

The establishment of systematic ABFE protocols facilitated their integration into practical research contexts, particularly within computer-aided drug design (CADD). Richard Friesner (born 1952), a computational chemist and co-founder of Schrödinger, Inc.—a leading developer of molecular modeling and drug discovery software—highlights the practical significance of ABFE methods for industrial applications:[xxxi] "ABFE calculations are of tremendous value in virtual screening. You take the top few thousand hits, run them through ABFE, and you get a better metric than any empirical scoring function.[xxxii] That's a place where theory made a significant contribution." Friesner's remarks reflect the broader adoption of rigorous physical models within high-throughput screening pipelines, where accurate prediction of binding affinities enables more effective prioritization of candidate molecules. Absolute free-energy methods offer detailed thermodynamic insights necessary for optimizing lead compounds, thereby supporting a more systematic and quantitative approach to ligand selection and refinement. However, despite the recognition granted to ABFE nowadays, the pioneering formal developments were often misunderstood and perceived negatively by the community. Publication of the foundational work on the calculation of ABFE encountered multiple difficulties,[xxxiii] and McCammon recalls that "The initial reception to this area of work was certainly surprisingly lukewarm at the best".[xxxiv]

### c) Transfer of Biomolecular Free-Energy Technology to the Industry

Alchemical free-energy methods such as TI and FEP were developed in academic settings beginning in the 1980s, with early applications by Tembe and McCammon,[21] Jorgensen and Ravimohan,[22] among others. During this period, academic researchers expressed substantial optimism that free-energy methodologies could eventually support a significant impact on pharmaceutical discovery. An early free-energy workshop was organized at CECAM in 1987, followed by additional meetings over the subsequent years. Scientific interest in free-energy calculations remained active throughout the 1988–1990 period. At the same time, the Alliant computer company, which designed and sold some of the earliest parallel computing systems [xxxv] organized several "free energy" workshops featuring prominent researchers in the field as part of its promotional efforts. [xxxvi] Despite this enthusiasm within academia, industrial engagement with alchemical free-energy techniques remained relatively cautious during this phase. In retrospect, the historical record suggests that large-scale industrial adoption of free-energy methods in the early 1990s was premature.

Some pharmaceutical research groups attempted to apply FEP methods,[124] but the computational platforms available at the time were cumbersome, and it was often infeasible to run simulations long enough to produce reliable results. Implementing such calculations required substantial expertise, including laborious system preparation, ligand parameterization, and management of long simulation times. Even when set up correctly, the computations could require several weeks to complete—a turnaround time incompatible with the fast-paced cycles of industrial drug development. Commenting on the trajectory of free-energy calculations and its impact on drug discovery research, McCammon says:[xxxvii]

> I think it's fair to say that FEP methods took a couple of decades to achieve the accuracy and speed needed to be adopted by industry. Earlier, there were indirect but important nudges, however. The new startup company Agouron Pharmaceuticals in La Jolla was excited by early FEP work. They recruited me to guide the startup of their computational work in 1987, and I wrote that part of their important NIH proposal and defended it in a site visit, leading to the start of the effort that developed the important HIV protease inhibitor nelfinavir (Viracept). FEP helped to get the grant, and was used in the company, but was not yet ready to help in practical work. But FEP has been used successfully in a number of companies in the current decade, most notably by the DE Shaw's company.[xxxvii]

The earliest application of free-energy methods to a substantial congeneric series of P38 ligands was published by David Pearlman in 2001 while at Vertex.[105] While the study represented an important methodological demonstration, the simulations required extensive computational resources and hundreds of picoseconds of sampling for each system, with total calculation times extending over weeks. Such timescales underscored the persistent gap between theoretical promise and industrial feasibility. David Pearlman had transitioned from Peter Kollman's research group to Vertex Pharmaceuticals in 1991.[xxiv] Founded in 1990, Vertex was among the first companies to articulate a model in which computational methods were intended to stand alongside experimental chemistry and biology as coequal contributors to drug discovery. Pearlman was joined by Govinda Rao, a former student of Chandra Singh who had contributed to the implementation of TI and FEP in AMBER. At Vertex, free-energy calculations were explored, though they did not yet constitute the primary computational approach to most problems. Across the industry more broadly, interest in free-energy simulations remained limited, reflecting the still modest role of computational chemistry within pharmaceutical research organizations. While a few large pharmaceutical companies maintained significant computational groups, many relied on small teams, and smaller firms often lacked dedicated computational personnel altogether.[xxiv]

The early development of free-energy methodologies had been characterized by substantial optimism about their future applicability to drug discovery. Academic pioneers such as McCammon, Jorgensen, Kollman, and Karplus made foundational contributions to alchemical free-energy methods, including TI and FEP. However, significant technical challenges persisted, and many critical issues remained unidentified during the early phases of methodological construction. Solutions to these problems would emerge only gradually over subsequent decades, including the formulation of soft-core shifted Lennard-Jones potentials,[125] long-range electrostatics,[126] statistical mechanical formalities about the treatment of dummy particles in single and dual topology,[98,108] reference-state definitions for absolute binding free energy,[111,116,117] enhanced sampling strategies for flexible side chains.[127,128] One major structural limitation during this period was the lack of generalized force fields for small molecules. Standard biomolecular force fields such as CHARMM, AMBER, GROMOS, and OPLS were originally parameterized for proteins and nucleic acids, not for drug-like compounds. This gap was gradually addressed by the development of all-atom OPLS,[129] MMFF (Merck Molecular Force Field),[130] GAFF (Generalized Amber Force Field),[131] and CGenFF (CHARMM Generalized Force Field).[132] But by the early 2000s, it was increasingly recognized that FEP calculations based on molecular dynamics simulations confronted severe practical limitations. The initial optimism of the 1980s and 1990s gave way to a phase of disenchantment, as unmet technical challenges tempered expectations. This trajectory bore similarities to the so-called "AI winter,"[xxxviii], when enthusiasm for artificial intelligence research collapsed temporarily under the weight of technical and operational barriers.

For much of the 1990s and early 2000s, skepticism toward the industrial viability of free-energy simulations persisted while continued, albeit slow, progress being made in academia. However, whereas academia valued free-energy calculations primarily as tools for theoretical investigation, the pharmaceutical industry required methods capable of delivering reliable predictions within commercially viable timescales. This divergence created a structural mismatch between methodological capabilities and industrial expectations. Friesner, co-founder of the molecular modeling software company Schrödinger, reflects on the barriers that delayed broader adoption. He notes:[xxxi]

> The adoption of FEP methods and explicit solvent simulations within the pharmaceutical industry experienced significant delays. While these techniques were crucial in academic research, they struggled to gain traction in the pharmaceutical sector for several decades. The foundations of these methodologies were established through the work of pioneering researchers, including the contributions of Jorgensen, Kollman, and Karplus and others, whose advancements were instrumental in the evolution of computational chemistry.[xxxi]

Jorgensen convincingly demonstrated the practical viability of alchemical FEP for computational lead discovery and optimization,[29] with calculations carried out using the Monte Carlo program MCPRO[133] with the all-atom OPLS force field for organic molecules.[129] Signaling an important first step toward an industrial implementation of FEP, MCPRO was adopted and marketed as MCPRO+ by Schrödinger in 2006. Friesner's account highlights that technical rigor alone was insufficient; operational reliability within high-throughput environments was the decisive standard. In industries driven by rapid iteration, theoretical soundness mattered only insofar as it translated into actionable, reproducible outcomes. He further elaborates on the quantitative thresholds that constrained early adoption:

> A fundamental challenge in applying computational methods to drug discovery was the required accuracy to rank molecules effectively. In the context of FEP, this accuracy was quantified by the root mean square (RMS) error, which needed to be close to 1 kcal/mol for the results to be meaningful and reliable for pharmaceutical applications. This threshold of 1 kcal/mol was essential for distinguishing between molecules in the synthesis queue.[23] Existing programs at the time were unable to achieve this level of accuracy, with errors in the range of 2–3 kcal/mol, which was insufficient for practical drug discovery applications. The difficulty in achieving an RMS error of 1 kcal/mol stemmed from multiple factors. Achieving such accuracy required not just precise computational techniques but also a comprehensive infrastructure to ensure that factors like protein preparation, protonation state assignment, and other preprocessing steps were handled appropriately. Without this integrated pipeline, the reliability of the FEP predictions would be compromised.[xxxi]

His observations highlight that predictive performance could not be achieved through algorithmic refinement alone. A reliable deployment of free-energy methods required the systematic integration of computational workflows, encompassing protein preparation, protonation state assignment, conformational sampling, and other pre-simulation steps. Without a cohesive infrastructural, free-energy calculations remained vulnerable to inconsistencies that undermined their being widely adopted within drug development environments.[23]

A decisive limitation during this period remained the computational capacity required to sufficiently sample the fluctuations of biomolecular systems; the timescales and sampling demands for robust free-energy predictions far exceeded the practical capabilities of available hardware.

Schrödinger's 2015 publication[7] arguably marked a pivotal step in making free-energy methods operational for industrial drug discovery. For the first time, FEP calculations demonstrated consistent predictive reliability within prospective, large-scale design campaigns. This achievement, nearly three decades after the earliest theoretical proposals, reflected not merely technical improvements, but a broader reconfiguration of methodological standards around operational reproducibility, robustness, and integration within pharmaceutical workflows. The recognition of this operational maturity invites reflection on the long developmental arc that preceded it. The 2015 landmark paper has been complemented by an extensive benchmark study, claiming to achieve accuracy (almost) comparable to experimental uncertainties between repeated measurements.[23] Simultaneously, while advanced free-energy methodologies are increasingly adopted by the industry, research in academia also contributes to demonstrate these are powerful tools for the discovery and optimization of novel drugs.[134,135] In particular, the work by Zhang et al[134] clearly demonstrates the critical role of FEP-guided lead optimization.

Reflecting on the three-decade development arc from early academic formulations to Schrödinger's publication of the FEP+ protocol, a key question arises: why did reliable industrial deployment materialize only in 2015, rather than earlier? Friesner suggests that this delay was not primarily due to conceptual shortcomings, but to the absence of sufficient computational and infrastructural support. He recalls: "Timing is everything. If I had been working on FEP in 1988, I might have made the same mistakes—underestimating the difficulties."[xxxi] Friesner's comment illustrates some of the key challenges: the alignment between methodological ambition and infrastructural feasibility remained elusive until the early 2010s. He describes how his perspective began to shift:

> Looking back, it wasn't until around 2010 that I really felt the computational infrastructure was ready for prime time. A lot of that had to do with going to talks and seeing how things were evolving. The big leap came with the advent of fast GPUs and the computer program Desmond, which made a huge difference. The cost-performance ratio improved dramatically, about a factor of ten in just a year. That was a real turning point, and it made me believe that it could work.[xxxi]

The technological maturation of hardware (through the emergence of GPUs) and software (e.g., Desmond) did not merely accelerate calculations; it reconfigured the operational expectations placed on computational methods. Improved cost-performance ratios allowed molecular dynamics simulations to reach timescales and sampling sufficiency that made rigorous prospective prediction feasible within pharmaceutical timelines.

At this juncture, free-energy calculations began to occupy a durable operational role within drug discovery pipelines. Friesner stresses that the slow trajectory toward industrial relevance was not anomalous, but rather emblematic of a broader historical pattern in the adoption of scientific innovations. As he notes: "This kind of skepticism isn't unique to FEP; it happens with many technologies. People always expect new technologies to revolutionize everything instantly. When they don't, there's a backlash. It's a pattern that repeats across history."[xxxi] He further illustrates this historical point humorously: "Newton published the *Principia* in 1687 but it took another 100 years to calculate the trajectory of a cannonball accurately".[xxxi] Friesner's reflection foregrounds that scientific methodologies often experience long periods of gestation, where early theoretical formulations must be patiently translated into practical, reproducible workflows through incremental technical and institutional adaptation. The transformation of free-energy calculations into robust industrial protocols was not the result of a single breakthrough, but of cumulative, distributed refinement across multiple layers of computational practice. The advent of fast and affordable GPU technology, while critical, should thus be seen as a catalyst rather than the sole cause. It provided the necessary conditions under which prior decades of theoretical innovation could finally realize operational maturity.

The historical trajectory of free-energy calculations in biomolecular systems also exemplifies the complex, often non-linear interactions between academic research and industrial application. Friesner emphasizes the indispensable role of early academic efforts in laying the theoretical groundwork that later enabled practical implementation. He remarks:

> Without all the groundwork done by people like Karplus, Jorgensen, and Kollman, we wouldn't have been able to make the breakthroughs we did. Their work laid the foundation. And, of course, you need someone to take that theoretical work and turn it into something practical, something that can be implemented in a product. If the academic groups hadn't existed, I don't think we ever would have done anything with FEP. I mean, we relied on all the fundamental work that they did. The truth is, none of us can take all the credit. It's been a massive collective effort. Everyone involved, from those who developed the theory to those who implemented it, made contributions that have brought us to where we are today. It's been one giant enterprise, and without those contributions, we probably wouldn't be here.[xxxi]

This testimony highlights the distributed character of innovation: industrial success in free-energy methods was not the outcome of isolated conceptual leaps, but the result of incremental stabilization across theoretical, methodological, and engineering domains. Frank Noé (born 1975), a computational chemist who moved from Berlin University to a research position at Microsoft, similarly reflects on the distinct but complementary roles of academia and industry: "Industry labs such as ours are really good at executing, i.e. large projects with a lot of planning and resources but a very narrow focus. On the other hand, academic labs are really good at exploring, i.e. they can quickly test out new ideas, let a 1000 flowers bloom and then pick the most promising ones. Academia develops and innovates ideas; industry then selects ideas and brings them to scale."[xxxix] Noé's remarks capture a broader dynamic wherein academia functions as an exploratory space for methodological proliferation, while industry acts as a selective and scaling apparatus. The eventual operationalization of free-energy methods thus depended on a long interplay between theoretical innovation, experimental trial, infrastructural development, and strategic industrial implementation.

While these acknowledgments underscore the foundational contributions of specific individuals and their groups, they do not fully capture the practical, infrastructural work that ultimately stabilized the field. The widespread use of free-energy methods rested on the iterative accumulation of operational expertise—including software refinement, parameter optimization, systematic benchmarking, and sustained efforts in documentation and training. Many protocols that came to define everyday practice were not formalized through flagship publications or centralized initiatives. Instead, they emerged through repeated local adaptation to specific architectures and simulation setups, alongside the informal circulation of problem-solving strategies across diverse research environments. The durability of these methods ultimately stemmed less from formal elegance than from consistent performance across heterogeneous conditions, including hardware constraints, molecular complexity, platform-specific dependencies, and evolving behaviors of simulation codebases. Viewed through this lens, the history of FEP and TI appears not as a chronicle of discrete conceptual breakthroughs, but as a record of how methodological reliability was progressively secured through the stabilization of computational practices, the recursive shaping of software by scientific need, and the embedding of simulation workflows within a diverse research culture.

## 7. NEW HORIZONS

Now that free-energy methods such as FEP and TI have been consolidated within many industrial workflows, now that they have gained "credibility", attention is shifting toward new modeling paradigms. This transition is not solely technical; it signals a reorganization of what counts as a credible modeling strategy, a transformation that is now unfolding across academic laboratories, industrial research groups, and distributed computational ecosystems. Nowadays, various commercial and open-source solutions for alchemical free-energy calculations continue to arise, an indication that the field continues to evolve. Multiple directions are gaining traction. One strand builds on the formal architecture of physics-based simulation, enhancing it through machine learning (ML). These efforts include the training of data-driven force fields tailored to quantum or empirical baselines, the use of neural networks to accelerate density functional approximations, and the deployment of generative models, notably diffusion-based architectures, to traverse configurational space more efficiently. Researchers have outlined frameworks where these elements are not simply adjuncts to traditional force fields but constitute reconfigurable units within modular simulation stacks, capable of being tuned or reassembled for different classes of molecular problems.

A second direction, emerging more visibly in data-rich contexts such as large pharmaceutical or materials science programs, shifts away from explicit simulation altogether. It turns instead to foundation models trained on broad molecular corpora to predict physical observables, including binding affinities, from structural descriptors alone. These models tend to suppress mechanistic transparency in favor of predictive yield. Their value lies less in explaining molecular interactions and more in delivering high-throughput screening capability that is decoupled from traditional simulation workflows, provided sufficient empirical data exist to support model generalization.

These developments do not imply a simplistic opposition between theory-driven and data-driven strategies; instead, they reflect the emergence of a pluralistic modeling landscape in which methods are evaluated according to context-dependent criteria: data availability, time constraints, interpretive requirements, and regulatory compatibility. Physics-based simulations are likely to remain essential to train AI methods where empirical coverage is sparse, where generalization across chemical space is required, or where cross-observable consistency imposes stricter theoretical constraints. In contrast, data-driven models may dominate in domains that prioritize speed, statistical reproducibility, or exploratory molecule generation at scale. The coexistence of these approaches recapitulates a historical logic familiar from the development of FEP and TI themselves. Alchemical free-energy methods gained "credibility" not through abstract theoretical formulation alone, but through an incremental process of refinement, standardization, testing, benchmarking, validation, and comparison with experiments, progressively achieving an alignment with the practical limitations of hardware technology and the real-world needs of the industry. This suggests that the evolution of emerging modeling paradigms will likely be shaped by similar forces before they gain their own credibility. It is also worth recalling that comparison of computations with experiments, again, will play an essential role. At the boundary between experiments and computations, modeling tools must continue to recalibrate themselves in response to shifting standards of reliability, interpretability, and practical utility.

## 8. CONCLUDING REMARKS

The historical trajectory of alchemical free-energy methods illustrates a transformation that cannot be described as a simple migration of formal statistical mechanics toward computational biology. Bridging the gap between mid-20th-century statistical mechanics and the mature workflows of the 2010s required more than the direct execution of ensemble-based statistical mechanical formalisms via computer simulations; it demanded an operational framework capable of managing the numerical, conceptual, and procedural complexities inherent in complex biomolecular transformations without clear physical analogs.

The earliest formulations of free-energy perturbation (Zwanzig, 1954) and thermodynamic integration (Kirkwood-derived expressions, late 1930s operationalized in the 1970s) provided mathematically rigorous pathways to compute free-energy differences. Yet these formalisms were designed within the closed theoretical constraints of ensemble averages and equilibrium statistical mechanics, without consideration for the algorithmic instantiation of these principles in sophisticated biomolecular systems. Practical implementations in the 1960s and 1970s focused on simple liquids and crystalline systems, where the operation of sampling configurational space was conceptually clear and relatively tractable.

The attempted migration of these methods into the biomolecular domain, following the earliest MD simulations of proteins in 1977, exposed limitations that were not merely computational but conceptual and structural. The concept of thermodynamic cycles, when applied to ligand binding, conformational change, or solvation processes, free-energy calculations involving biomolecules required introducing non-physical intermediate states, artificial coupling parameters, and mathematical constructs such as dummy atoms. These interventions were necessary to interpolate between chemically distinct states but introduced profound numerical instabilities—particularly at the end points of transformations where singularities, poorly behaved gradients, and discontinuities in force-space became sources of failure.

By the mid-1980s, the main conceptual ideas introduced by McCammon (1984) and Jorgensen (1985) and subsequently implemented by Kollman (1987) and Karplus (1989) provided a scaffolding to decompose chemically meaningful free-energy differences into computationally manageable paths. But the existence of this formal decomposition did not resolve the underlying challenges. The methods suffered from poor phase-space overlap between neighboring states, systematic breakdowns in sampling convergence, as well as sensitivity to force-field parameters that were themselves underdetermined for small molecules and heterogeneous biomolecular systems. The initial optimism surrounding alchemical free-energy methods was shaped by the belief that rational drug design could be transformed through direct, predictive modeling of molecular binding affinities. However, significant challenges soon tempered expectations.

Between 1985 and 2005, efforts to stabilize alchemical free-energy workflows were distributed across several fronts. Theoretical reformulation together with a series of algorithmic strategies were introduced to manage the persistent instabilities and breakdowns inherent in alchemical transformations. These included soft-core potentials to regularize Lennard-Jones singularities; dual-topology frameworks to handle topological discontinuities; modified treatments of long-range electrostatics embedded directly within simulation software; positional restraint schemes to prevent unphysical drift during ligand transformations; and endpoint-specific corrections to manage Coulombic singularities during electrostatic decoupling. At the same time, the absence of generalized force fields explicitly parameterized for small drug-like molecules posed a structural barrier to the construction of robust objective workflows. While this gap was progressively addressed, the stabilization of free-energy methods could not be achieved through algorithmic modifications alone. It required the codification of simulation protocols into repeatable procedures; the careful management of boundary conditions to control artifacts from periodicity and electrostatics; the formal definition of convergence thresholds to determine when sampling was adequate; the systematic quantification of statistical uncertainty through error estimation methods; and the alignment of computational workflows with the time-sensitive demands of drug discovery. In pharmaceutical contexts, theoretical accuracy was necessary but not sufficient. Free-energy estimates also had to deliver operational reproducibility, robustness across software environments, and compatibility with the procedural demands of drug discovery (e.g., fixed timelines for lead optimization, decision-gating milestones, and the requirement for auditable, reproducible computational outputs).

By the 2010s, the stabilization of free-energy methods no longer rested on isolated algorithmic improvements. It reflected the entanglement of several mutually dependent processes: the consolidation of software infrastructures that embedded specific algorithmic choices and boundary conditions; the formalization of validation pipelines involving retrospective and prospective benchmarks; the integration of sampling-enhancement techniques to address the timescale separation problem in biomolecular systems; and the increasing alignment of simulation outputs with industrial decision-making processes, including lead optimization in medicinal chemistry.

Emerging from this reconstruction is the observation that the operational reliability of free-energy methods cannot be attributed to the intrinsic properties of statistical mechanical formalisms. Nor can it be reduced to computational brute force. Instead, reliability was assembled through the long-term calibration of mathematical approximations, force-field parameterization strategies, algorithmic conventions, error-tolerance thresholds, and the collaborative engineering of software environments that encoded, stabilized, and constrained these methodological choices. Formal derivations remained necessary, but they were never sufficient. What made free-energy methods usable—first in academic settings and eventually in industrial workflows—was the domestication of their inherent numerical fragility into practices that bounded instability, defined acceptable uncertainty, and embedded these limits into decision-relevant computational procedures.

The adoption of biomolecular free-energy methods in the industry is not a story of theoretical finality. It is a history of stabilization under operational constraint—where reliability was progressively achieved by gradually managing the uncertainty rather than its elimination. Trust in FEP calculations did not flow automatically from formal derivability; it had to be earned through testing, validation, and the crafting of procedures that made uncertainty tractable. What became standardized was less a universal truth about molecular systems than a set of procedural compromises, technical heuristics, and software architectures that rendered previously intractable calculations operational within the evolving "epistemic economy" of molecular simulations. Within this context, the specific and contingent conditions under which knowledge claims are made actionable—through software, benchmarks, procedural routines, and negotiated tolerances. The consolidation of free-energy workflows reflects a recursive system of mutual dependencies, in which mathematical rigor, software engineering, parameter standardization, hardware acceleration, and collaborative validation co-developed as interdependent components. Each was continuously shaped, constrained, and adjusted in response to the others, through recursive alignment governed by the technical, organizational, and procedural constraints of computational molecular science. At the same time, this history reflects how simulation itself became structurally embedded in the industrial workflow of pharmaceutical research, where computational predictions were progressively integrated into decision-making, project management, and the procedural architecture of drug design.

## Acknowledgements

All interviews conducted for this study are available from the corresponding author upon request. We are grateful to the interviewees for their willingness to share their recollections and for allowing us to use them in this work. The authors are grateful to Martin Karplus, William L. Jorgensen, and J. Andrew McCammon for their generosity in sharing recollections and clarifying technical details, as well as for providing feedback on specific aspects of the historical reconstruction. They also thank Arieh Warshel, Stefan Boresch, Dave Case, Piotr Cieplak, Richard A. Friesner, Jiali Gao, Michael Gilson, Marci Karplus, Mihaly Mezei, Frank Noé, David Pearlman, Ed Samulski, Tjerk Straatsma, Bruce Tidor, and Wilfred F. van Gunsteren for their cooperation and for supplying clarifications, documents, or archival material. Minor language refinements and assistance in organizing interview transcripts and textual material were provided by advanced editing tools. The authors are deeply indebted to William Jorgensen for generously sharing numerous photographs and documents from his personal archives, and to Marci Karplus (spouse of Martin Karplus), Frederieke Berendsen (daughter of Herman Berendsen), Eli Kollman (son of Peter Kollman), Ken Alder (son of Bernie Alder), and Robert Bonner (nephew of John Kirkwood) for kindly granting permission to use the photographs. This work was supported by the Beijing Natural Science Foundation (北京市自然科学基金委员会) under Grant IS23131 (D.M.) and by the National Institute of Health via grant R35-GM152124 and the National Science Foundation via grant MCB-2309048 (B.R.).

## Biographies

**Benoît Roux** was born in Montréal, Canada, in 1958. In 1981, he received a B.Sc. in Physics from the University of Montréal, followed by a M.Sc. in Biophysics in 1985 under the supervision of Rémy Sauvé. In 1990, he obtained a Ph.D. in Biophysics from Harvard University under the direction of Martin Karplus. In the last decade, he has held positions at the University of Montréal and the Weill Medical College of Cornell University. Since 2005, he is Professor in the Department of Biochemistry and Molecular Biology at the University of Chicago with a joint appointment as Senior Computational Biologist at Argonne National Laboratory.

**Daniele Macuglia** was born in Tolmezzo, Italy, in 1984. He graduated in Physics from the University of Pavia and the Scuola Universitaria Superiore IUSS, and earned his Ph.D. in Conceptual and Historical Studies of Science at the University of Chicago in 2017, supervised during his studies by Robert Richards and Leo Kadanoff. He later served as a research fellow at the Stevanovich Institute on the Formation of Knowledge and the Neubauer Collegium. Since 2021 he has been Assistant Professor in the Department of History of Science, Technology, and Medicine at Peking University. His research centers on the history of computational statistical mechanics and molecular simulation.

**Giovanni Ciccotti** was born in Rome, Italy, in 1943. He earned a MS in Physics in 1967, at Rome University "La Sapienza" in the field of theoretical elementary particles. He held the position of Professor of Structure of Matter in the Physics Department until 2014 and has been Emeritus Professor at "La Sapienza" since 2015. His activity in molecular dynamics and computational statistical mechanics started in Paris in 1974, while at the *Centre Européen de Calcul Atomique et Moléculaire* (CECAM). He received the Berni Alder prize from CECAM in 1999 for his pioneering contributions to molecular dynamics, including the non-equilibrium subtraction technique to study transport phenomena, and constrained molecular dynamics.

## References

1. Helmholtz, H. v. Über die Thermodynamik der chemischen Vorgänge. *Sitzungsberichte Der Königlich Preussischen Akademie Der Wissenschaften Zu Berlin* **1**, 22–39, (1882).
2. Helmholtz, H. v. Über die Thermodynamik der chemischen Vorgänge. Zweiter Beitrag. Versuche an Chlorzink-Kalomel-Elementen. *Sitzungsberichte Der Königlich Preussischen Akademie Der Wissenschaften Zu Berlin* **2**, 825-836, (1882).
3. Helmholtz, H. v. Über die Thermodynamik der chemischen Vorgänge. Dritter Beitrag. *Sitzungsberichte Der Königlich Preussischen Akademie Der Wissenschaften Zu Berlin* 647–661 (1883).
4. Gibbs, J. W. On the equilibrium of heterogeneous substances. . *Transactions of the Connecticut Academy of Arts and Sciences* **3**, 108–248 (1876).
5. Gibbs, J. W. On the equilibrium of heterogeneous substances. . *Transactions of the Connecticut Academy of Arts and Sciences* **3**, 343–524, (1878).
6. Gibbs., J. W. *Elementary Principles in Statistical Mechanics Developed with Especial Reference to the Rational Foundation of Thermodynamics*. (Charles Scribner's Sons, 1902).
7. Wang, L. *et al.* Accurate and reliable prediction of relative ligand binding potency in prospective drug discovery by way of a modern free-energy calculation protocol and force field. *J Am Chem Soc* **137**, 2695-2703, (2015).
8. Metropolis, N., Rosenbluth, A. W., Rosenbluth, M. N., Teller, A. H. & Teller, E. Equation of State Calculations by Fast Computing Machines. *J Chem Phys* **21**, 1087-1092, (1953).
9. Alder, B. J. & Wainwright, T. E. Phase Transition for a Hard Sphere System. *J Chem Phys* **27**, 1208-1209, (1957).
10. Rahman, A. Correlations in motion of atoms in liquid argon. *Phys. Rev.* **136**, A405-A411, (1964).
11. Stillinger, F. H. & Rahman, A. Improved simulation of liquid water by molecular dynamics. *J. Chem. Phys.* **60**, 1545-1557, (1974).
12. Mcdonald, I. R. & Singer, K. Calculation of Thermodynamic Properties of Liquid Argon from Lennard-Jones Parameters by a Monte Carlo Method. *Discuss Faraday Soc* 40-&, (1967).
13. Mcdonald, I. R. & Singer, K. Machine Calculation of Thermodynamic Properties of a Simple Fluid at Supercritical Temperatures. *J Chem Phys* **47**, 4766-&, (1967).
14. Wood, W. W. & Parker, F. R. Monte Carlo Equation of State of Molecules Interacting with the Lennard-Jones Potential .1. Supercritical Isotherm at About Twice the Critical Temperature. *J Chem Phys* **27**, 720-733, (1957).
15. Hansen, J. P. & Verlet, L. Phase Transitions of Lennard-Jones System. *Phys Rev* **184**, 151-&, (1969).
16. Valleau, J. P. & Card, D. N. Monte-Carlo Estimation of Free-Energy by Multistage Sampling. *J Chem Phys* **57**, 5457-&, (1972).
17. Patey, G. N. & Valleau, J. P. Free-Energy of Spheres with Dipoles - Monte-Carlo with Multistage Sampling. *Chem Phys Lett* **21**, 297-300, (1973).
18. Bennett, C. H. Efficient estimation of free-energy differences from Monte-Carlo data. *J. Comp. Chem.* **22**, 245-268, (1976).
19. McCammon, J. A., Gelin, B. R. & Karplus, M. Dynamics of folded proteins. *Nature* **267**, 585-590, (1977).
20. Macuglia, D., Roux, B. & Ciccotti, G. The emergence of protein dynamics simulations: how computational statistical mechanics met biochemistry. *Eur Phys J H* **47**, (2022).
21. Tembe, B. L. & Mccammon, J. A. Ligand Receptor Interactions. *Comput Chem* **8**, 281-283, (1984).
22. Jorgensen, W. L. & Ravimohan, C. Monte-Carlo Simulation of Differences in Free-Energies of Hydration. *J Chem Phys* **83**, 3050-3054, (1985).
23. Ross, G. A., Lu, C., Scarabelli, G., Albanese, S. K., Houang, E., Abel, R., Harder, E. D. & Wang, L. The maximal and current accuracy of rigorous protein-ligand binding free energy calculations. *Commun Chem* **6**, 222, (2023).
24. van Gunsteren, W. F. & Berendsen, H. J. Thermodynamic cycle integration by computer simulation as a tool for obtaining free energy differences in molecular chemistry. *J Comput Aided Mol Des* **1**, 171-176, (1987).

25. Beveridge, D. L. & Dicapua, F. M. Free energy via molecular simulation - applications to chemical and biomolecular systems. *Annu. Rev. Biophys. Bio.* **18**, 431-492, (1989).
26. Kollman, P. A. Free energy calculations: applications to chemical and biochemical phenomena. *Chem. Rev.* **93**, 2395-2417, (1993).
27. Simonson, T., Archontis, G. & Karplus, M. Free energy simulations come of age: protein-ligand recognition. *Acc Chem Res* **35**, 430-437, (2002).
28. Chipot, C. & Pohorille, A. *Free Energy Calculations: Theory and Applications in Chemistry and Biology.* Vol. 86 (Springer, 2007).
29. Jorgensen, W. L. Efficient drug lead discovery and optimization. *Accounts of chemical research* **42**, 724-733, (2009).
30. Deng, Y. & Roux, B. Computations of Standard Binding Free Energies with Molecular Dynamics Simulations. *J Phys Chem B* **113**, 2234–2246, (2009).
31. Mobley, D. L. & Dill, K. A. Binding of small-molecule ligands to proteins: "what you see" is not always "what you get". *Structure* **17**, 489-498, (2009). **PMC2756098**
32. Born, M. Volumen und Hydratationswarme der Ionen. *Z. Phys.* **1**, 45-48, (1920).
33. Roux, B., Yu, H. A. & Karplus, M. Molecular-Basis for the Born Model of Ion Solvation. *J Phys Chem* **94**, 4683-4688, (1990).
34. Debye, P. & Hückel, E. Zur Theorie der Elektrolyte. II. Das Grenzgesetz für die elktrische Leitfahigkeit. *Phys. Z.* **24**, 305-325, (1923).
35. Kirkwood, J. G. Statistical Mechanics of Fluid Mixtures. *J. Chem. Phys.* **3**, 300-313, (1935).
36. Berendsen, H. J. Biophysical applications of molecular dynamics. *Computer Physics Communications* **44**, 233—242, (1985).
37. Berendsen, H. J. Dynamic simulation as an essential tool in molecular modeling. (1988).
38. Fleischman, S. H. & Brooks III, C. L. Thermodynamics of aqueous solvation: Solution properties of alcohols and alkanes. *J. Chem. Phys.* **87**, 3029-3037, (1987).
39. Gao, J., Kuczera, K., Tidor, B. & Karplus, M. Hidden thermodynamics of mutant proteins: A molecular dynamics study. *Science* **244**, 1069-1072, (1989).
40. Karplus, M. & Petsko, G. A. Molecular dynamics simulations in biology. *Nature* **347**, 631-639, (1990).
41. Alder, B. & Wainwright, T. Studies in molecular dynamics .1. General method. *J. Chem. Phys.* **31**, 459-466, (1959).
42. Wood, W. W. & Jacobson, J. D. Preliminary Results from a Recalculation of the Monte Carlo Equation of State of Hard Spheres. *J Chem Phys* **27**, 1207-1208, (1957).
43. Tidor, B. Simulated annealing on free energy surfaces by a combined molecular dynamics and Monte Carlo approach. *The Journal of Physical Chemistry* **97**, 1069-1073, (1993).
44. Kong, X. & Brooks, C. L. λ-dynamics: A new approach to free energy calculations. *The Journal of chemical physics* **105**, 2414-2423, (1996).
45. Raman, E. P., Paul, T. J., Hayes, R. L. & Brooks III, C. L. Automated, accurate, and scalable relative protein–ligand binding free-energy calculations using lambda dynamics. *Journal of chemical theory and computation* **16**, 7895-7914, (2020).
46. Zwanzig, R. W. High temperature equation of state by a perturbation method. *J. Chem. Phys.* **22**, 1420-1426, (1954).
47. Landau, L. D. & Lifshitz, E. M. *Course of Theoretical Physics, Vol. 5* (Pergamon Press Ltd., 1958).
48. Jorgensen, W. L. & Thomas, L. L. Perspective on Free-Energy Perturbation Calculations for Chemical Equilibria. *J Chem Theory Comput* **4**, 869-876, (2008).
49. Torrie, G. M. & Valleau, J. P. Monte Carlo Free Energy Estimates Using Non-Boltzmann Sampling: Application to the Sub-Critical Lennard-Jones Fluid. *Chem. Phys. Lett.* **28**, 578-581, (1974).
50. Torrie, G. M. & Valleau, J. P. Nonphysical sampling distribution in Monte Carlo free-energy estimation:umbrella sampling. *J.Comp.Phys.* **23**, 187-199, (1977).
51. Sugita, Y. & Okamoto, Y. Replica-exchange molecular dynamics method for protein folding. *Chem Phys Lett* **314**, 141-151, (1999).
52. Laio, A. & Parrinello, M. Escaping free-energy minima. *Proc Natl Acad Sci U S A* **99**, 12562-12566, (2002).
53. Widom, B. Some Topics in Theory of Fluids. *J Chem Phys* **39**, 2808-&, (1963).
54. Michels, A., Wijker, H. & Wijker, H. K. Isotherms of Argon between O-Degrees-C and 150-Degrees-C and Pressures up to 2900 Atmospheres. *Physica* **15**, 627-633, (1949).
55. Michels, A., Levelt, J. M. & Wolkers, G. J. Thermodynamic Properties of Argon at Temperatures between O-Degree-C and 140-Degree-C and at Densities up to 640 Amagat (Pressures up to 1050 Atm). *Physica* **24**, 769-794, (1958).
56. Hoover, W. G. & Ree, F. H. Use of Computer Experiments to Locate Melting Transition and Calculate Entropy in Solid Phase. *J Chem Phys* **47**, 4873-+, (1967).
57. Patey, G. & Valleau, J. Dipolar hard spheres: A Monte Carlo study. *J Chem Phys* **61**, 534-540, (1974).
58. Ferrenberg, A. M. & Swendsen, R. H. Optimized Monte-Carlo Data-Analysis. *Phys. Rev. Lett.* **63**, 1195-1198, (1989).
59. Kumar, S., Bouzida, D., Swendsen, R. H., Kollman, P. A. & Rosenberg, J. M. The Weighted Histogram Analysis Method for free-energy calculations on biomolecules. I. The method. *J. Comp. Chem.* **13**, 1011-1021, (1992).
60. Mezei, M., Swaminathan, S. & Beveridge, D. L. Abinitio Calculation of Free-Energy of Liquid Water. *J Am Chem Soc* **100**, 3255-3256, (1978).
61. Postma, J. P. M., Berendsen, H. J. C. & Haak, J. R. Thermodynamics of Cavity Formation in Water - a Molecular-Dynamics Study. *Faraday Symp Chem S*, 55-67, (1982).
62. Warshel, A. Dynamics of Reactions in Polar-Solvents - Semi-Classical Trajectory Studies of Electron-Transfer and Proton-Transfer Reactions. *J Phys Chem-Us* **86**, 2218-2224, (1982).
63. Warshel, A. in *Specificity in Biological Interactions (Working Group at the Pontifical Academy of Sciences November 9–11, 1983).* (ed C. Chagas, Pullman, B. (eds) ) 59-81 (Dordrecht: Springer 1984).
64. Helfand, E., Reiss, H., Frisch, H. L. & Lebowitz, J. L. Scaled Particle Theory of Fluids. *J Chem Phys* **33**, 1379-1385, (1960).
65. Warshel, A. & Sussman, F. Toward computer-aided site-directed mutagenesis of enzymes. *Proceedings of the National Academy of Sciences of the United States of America* **83**, 3806-3810, (1986).
66. Hwang, J. K. & Warshel, A. Semiquantitative calculations of catalytic free energies in genetically modified enzymes. *Biochemistry* **26**, 2669-2673, (1987).
67. Lybrand, T. P., McCammon, J. A. & Wipff, G. Theoretical calculation of relative binding affinity in host-guest systems. *Proc Natl Acad Sci USA* **83**, 833-835, (1986). **322963**
68. Wong, C. F. & Mccammon, J. A. Dynamics and Design of Enzymes and Inhibitors. *J Am Chem Soc* **108**, 3830-3832, (1986).
69. McCammon, J. A. Autobiography of J. Andrew McCammon. *J Phys Chem B* **120**, 8057-8060, (2016).
70. Jorgensen, W. L., Buckner, J. K., Boudon, S. & Tirado-Rives, J. Efficient computation of absolute free energies of binding by computer simulations. Application to methane dimer in water. *J. Chem. Phys.* **89**, 3742-3746, (1988).
71. Jorgensen, W. L. Interactions between Amides in Solution and the Thermodynamics of Weak Binding. *J Am Chem Soc* **111**, 3770-3771, (1989).
72. Battimelli, G., Ciccotti, G. & Greco, P. *Computer Meets Theoretical Physics: The New Frontier of Molecular Simulation.* (Cham: Springer Nature, 2020).

73. McQuarrie, D. A. *Statistical Mechanics*. (Harper and Row, 1976).
74. Gunsteren, W. F. v. & Berendsen, H. J. C. *The GROMOS Software for (Bio)Molecular Simulation*, <https://www.gromos.net/gromos87/GROMOS87_manual.pdf> (1987).
75. Lybrand, T. P., Ghosh, I. & McCammon, J. A. Hydration of Chloride and Bromide Anions: Determination of Relative Free Energy by Computer Simulation. *J. Am. Chem. Soc.* **107**, 7793-7794, (1985).
76. Bennett, C. H. Mass tensor molecular dynamics. *Journal of Computational Physics* **19**, 267-279, (1975).
77. Hopkins, C. W., Le Grand, S., Walker, R. C. & Roitberg, A. E. Long-Time-Step Molecular Dynamics through Hydrogen Mass Repartitioning. *J Chem Theory Comput* **11**, 1864-1874, (2015).
78. Weiner, P. K. & Kollman, P. A. Amber - Assisted Model-Building with Energy Refinement - a General Program for Modeling Molecules and Their Interactions. *J Comput Chem* **2**, 287-303, (1981).
79. Bash, P. A., Singh, U. C., Langridge, R. & Kollman, P. A. Free energy calculations by computer simulation. *Science* **236**, 564-568, (1987).
80. Bash, P. A., Singh, U. C., Brown, F. K., Langridge, R. & Kollman, P. A. Calculation of the relative change in binding free energy of a protein-inhibitor complex. *Science* **235**, 574-576, (1987).
81. Rao, S. N., Singh, U. C., Bash, P. A. & Kollman, P. A. Free energy perturbation calculations on binding and catalysis after mutating Asn 155 in subtilisin. *Nature* **328**, 551-554, (1987).
82. Brooks, B. R., Bruccoleri, R. E., Olafson, B. D., States, D. J., Swaminathan, S. & Karplus, M. CHARMM: A program for macromolecular energy minimization and dynamics calculations. *J. Comput. Chem.* **4**, 187-217, (1983).
83. Brooks, B. R. *et al.* CHARMM: the biomolecular simulation program. *J Comput Chem* **30**, 1545-1614, (2009).
84. Hwang, W. *et al.* CHARMM at 45: Enhancements in Accessibility, Functionality, and Speed. *The journal of physical chemistry. B* **128**, 9976-10042, (2024).
85. Karplus, M. *Spinach on the ceiling: the multifaceted life of a theoretical chemist.* (World Scientific Publishing Europe Ltd., 2020).
86. Field, M. J., Bash, P. A. & Karplus, M. A combined Quantum Mechanical and molecular mechanical potential for molecular dynamics simulations. *J. Comp. Chem.* **11**, 700-733, (1990).
87. Bash, P. A., Field, M. J. & Karplus, M. Free-Energy Perturbation Method for Chemical-Reactions in the Condensed Phase - a Dynamical-Approach Based on a Combined Quantum and Molecular Mechanics Potential. *J Am Chem Soc* **109**, 8092-8094, (1987).
88. Straatsma, T. P. & Mccammon, J. A. Argos, a Vectorized General Molecular-Dynamics Program. *J Comput Chem* **11**, 943-951, (1990).
89. Aprà, E. & E. J. Bylaska, W. A. d. J., N. Govind, K. Kowalski, T. P. Straatsma, M. Valiev, H. J. J. van Dam, Y. Alexeev, J. Anchell, V. Anisimov, F. W. Aquino, R. Atta-Fynn, J. Autschbach, N. P. Bauman, J. C. Becca, D. E. Bernholdt, K. Bhaskaran-Nair, S. Bogatko, P. Borowski, J. Boschen, J. Brabec, A. Bruner, E. Cauët, Y. Chen, G. N. Chuev, C. J. Cramer, J. Daily, M. J. O. Deegan, T. H. Dunning Jr., M. Dupuis, K. G. Dyall, G. I. Fann, S. A. Fischer, A. Fonari, H. Früchtl, L. Gagliardi, J. Garza, N. Gawande, S. Ghosh, K. Glaesemann, A. W. Götz, J. Hammond, V. Helms, E. D. Hermes, K. Hirao, S. Hirata, M. Jacquelin, L. Jensen, B. G. Johnson, H. Jónsson, R. A. Kendall, M. Klemm, R. Kobayashi, V. Konkov, S. Krishnamoorthy, M. Krishnan, Z. Lin, R. D. Lins, R. J. Littlefield, A. J. Logsdail, K. Lopata, W. Ma, A. V. Marenich, J. Martin del Campo, D. Mejia-Rodriguez, J. E. Moore, J. M. Mullin, T. Nakajima, D. R. Nascimento, J. A. Nichols, P. J. Nichols, J. Nieplocha, A. Otero-de-la-Roza, B. Palmer, A. Panyala, T. Pirojsirikul, B. Peng, R. Peverati, J. Pittner, L. Pollack, R. M. Richard, P. Sadayappan, G. C. Schatz, W. A. Shelton, D. W. Silverstein, D. M. A. Smith, T. A. Soares, D. Song, M. Swart, H. L. Taylor, G. S. Thomas, V. Tipparaju, D. G. Truhlar, K. Tsemekhman, T. Van Voorhis, Á. Vázquez-Mayagoitia, P. Verma, O. Villa, A. Vishnu, K. D. Vogiatzis, D. Wang, J. H. Weare, M. J. Williamson, T. L. Windus, K. Woliński, A. T. Wong, Q. Wu, C. Yang, Q. Yu, M. Zacharias, Z. Zhang, Y. Zhao, and R. J. Harrison. NWChem: Past, present, and future. *J. Chem. Phys.* **152**, 184102, (2020).
90. Straatsma, T. P., Berendsen, H. J. C. & Postma, J. P. M. Free-energy of hydrophobic hydration - a molecular-dynamics study of noble-gases in water *J. Chem. Phys.* **85**, 6720-6727, (1986).
91. Mark, A. E. & van Gunsteren, W. F. Decomposition of the free energy of a system in terms of specific interactions. Implications for theoretical and experimental studies. *Journal of molecular biology* **240**, 167-176, (1994).
92. Smith, P. E. & van Gunsteren, W. F. When are free energy components meaningful? *The Journal of Physical Chemistry* **98**, 13735-13740, (1994).
93. Pearlman, D. A. & Kollman, P. A. The overlooked bond-stretching contribution in free energy perturbation calculations. *The Journal of chemical physics* **94**, 4532-4545, (1991).
94. Phillips, J. C., Braun, R., Wang, W., Gumbart, J., Tajkhorshid, E., Villa, E., Chipot, C., Skeel, R. D., Kale, L. & Schulten, K. Scalable molecular dynamics with NAMD. *J. Comput. Chem.* **26**, 1781-1802, (2005).
95. Case, D. A., Cheatham III, T. E., Darden, T., Gohlke, H., Luo, R., Merz Jr, K. M., Onufriev, A., Simmerling, C., Wang, B. & Woods, R. J. The Amber biomolecular simulation programs. *J Comput Chem* **26**, 1668-1688, (2005).
96. Boresch, S. *Methodological Studies of Free Energy Simulations* Ph.D. thesis, Harvard University (1997).
97. Boresch, S. & Karplus, M. The Jacobian factor in free energy simulations. *J Chem Phys* **105**, 5145-5154, (1996).
98. Boresch, S. & M.Karplus. The role of bonded terms in free energy simulations: Theoretical analysis *J Phys Chem A* **103**, 103-118, (2000).
99. Boresch, S. & M.Karplus. The role of bonded terms in free energy simulations. 2. Calculation of their influence on free energy differences of solvation. *J Phys Chem A* **103**, 119-136, (2000).
100. Prevost, M., Wodak, S. J., Tidor, B. & Karplus, M. Contribution of the hydrophobic effect to protein stability: analysis based on simulations of the Ile-96----Ala mutation in barnase. *Proc Natl Acad Sci U S A* **88**, 10880-10884, (1991).
101. Tidor, B. & Karplus, M. Simulation analysis of the stability mutant R96H of T4 lysozyme. *Biochemistry* **30**, 3217-3228, (1991).
102. Boresch, S., Archontis, G. & M.Karplus. Free Energy Simulations: The Meaning of The Individual Contributions from a Component Analysis Proteins. *Proteins* **20**, 25-33, (1994).
103. Straatsma, T. P., Zacharias, M. & Mccammon, J. A. Holonomic Constraint Contributions to Free-Energy Differences from Thermodynamic Integration Molecular-Dynamics Simulations. *Chem Phys Lett* **196**, 297-302, (1992).
104. Wang, L. & Hermans, J. Change of Bond-Length in Free-Energy Simulations - Algorithmic Improvements, but When Is It Necessary. *J Chem Phys* **100**, 9129-9139, (1994).
105. Pearlman, D. A. & Charifson, P. S. Are free energy calculations useful in practice? A comparison with rapid scoring functions for the p38 MAP kinase protein system. *J Med Chem* **44**, 3417-3423, (2001).
106. Herschbach, D. R., Johnston, H. S. & Rapp, D. Molecular Partition Functions in Terms of Local Properties. *J Chem Phys* **31**, 1652-1661, (1959).
107. Donnan, F. G. & Haas, A. *Commentary on the Scientific Writings of J. Willard Gibbs*. (Yale University Press, 1936).

108. Shobana, S., Roux, B. & Andersen, O. S. Free energy simulations: Thermodynamic reversibility and variability. *Journal of Physical Chemistry B* **104**, 5179-5190, (2000).
109. Fleck, M., Wieder, M. & Boresch, S. Dummy Atoms in Alchemical Free Energy Calculations. *Journal of chemical theory and computation* **17**, 4403-4419, (2021).
110. Hermans, J. & Shankar, S. The free energy of xenon binding to myoglobin from molecular dynamics simulation. *Isr J Chem* **27**, 225-227, (1986).
111. Hermans, J. & Wang, L. Inclusion of Loss of Translational and Rotational Freedom in Theoretical Estimates of Free Energies of Binding. Application to a Complex of Benzene and Mutant T4 Lysozyme. *J. Am. Chem. Soc.* **119**, 2707-2714, (1997).
112. Hermans, J., Berendsen, H. J. C., Gunsteren, W. F. v. & Postma, J. P. M. A consistent empirical potential for water-protein interactions. *Biopol* **23**, 1513-1518, (1984).
113. Cieplak, P. & Kollman, P. A. Calculation of the Free-Energy of Association of Nucleic-Acid Bases in Vacuo and Water Solution. *J Am Chem Soc* **110**, 3734-3739, (1988).
114. Wade, R. & McCammon, J. A. A molecular dynamics study of thermodynamic and structural aspects of the hydration of cavities in proteins. *Biopol.* **31**, 919-931, (1991).
115. Woo, H. J. & Roux, B. Calculation of absolute protein-ligand binding free energy from computer simulations. *Proc. Natl. Acad. Sci. U.S.A.* **102**, 6825-6830, (2005).
116. Roux, B., Nina, M., Pomès, R. & Smith, J. C. Thermodynamic stability of water molecules in the bacteriorhodopsin proton channel: A molecular dynamics free energy perturbation study. *Biophys J* **71**, 670-681, (1996).
117. Gilson, M. K., Given, J. A., Bush, B. L. & McCammon, J. A. The statistical-thermodynamic basis for computation of binding affinities: A critical review. *Biophys. J.* **72**, 1047-1069, (1997).
118. Boresch, S., Tettinger, F., Leitgeb, M. & Karplus, M. Absolute binding free energies: A quantitative approach for their calculation. *J. Phys. Chem. B* **107**, 9535-9551, (2003).
119. Deng, Y. Q. & Roux, B. Calculation of standard binding free energies: Aromatic molecules in the T4 lysozyme L99A mutant. *J. Chem. Theo. Comp.* **2**, 1255-1273, (2006).
120. Wang, J. Y., Deng, Y. Q. & Roux, B. Absolute binding free energy calculations using molecular dynamics simulations with restraining potentials. *Biophysical Journal* **91**, 2798-2814, (2006).
121. Fu, H., Chen, H., Blazhynska, M., Goulard Coderc de Lacam, E., Szczepaniak, F., Pavlova, A., Shao, X., Gumbart, J. C., Dehez, F., Roux, B., Cai, W. & Chipot, C. Accurate determination of protein:ligand standard binding free energies from molecular dynamics simulations. *Nat Protoc* **17**, 1114-1141, (2022).
122. Aqvist, J., Medina, C. & Samuelsson, J. E. A new method for predicting binding affinity in computer-aided drug design. *Protein Eng* **7**, 385-391, (1994).
123. Roux, B. & Chipot, C. Editorial Guidelines for Computational Studies of Ligand Binding Using MM/PBSA and MM/GBSA Approximations Wisely. *J Phys Chem B* **128**, 12027-12029, (2024).
124. Lilly, E. *Implementing the Cray 2 at Eli Lilly*, <https://youtu.be/ljw03V3nbF0?si=kTFw1HhLRBDCJ4Lt> (1991).
125. Zacharias, M., Straatsma, T. P. & Mccammon, J. A. Separation-Shifted Scaling, a New Scaling Method for Lennard-Jones Interactions in Thermodynamic Integration. *J Chem Phys* **100**, 9025-9031, (1994).
126. Darden, T. A., Toukmaji, A. & Pedersen, L. G. Long-range electrostatic effects in biomolecular simulations. *Journal de Chimie Physique et de Physico-Chimie Biologique* **94**, 1346-1364, (1997).
127. Jiang, W. & Roux, B. Free Energy Perturbation Hamiltonian Replica-Exchange Molecular Dynamics (FEP/H-REMD) for Absolute Ligand Binding Free Energy Calculations. *J. Chem. Theo. Comp.* **6**, 2559-2565, (2010).
128. Wang, L., Friesner, R. A. & Berne, B. J. Replica exchange with solute scaling: a more efficient version of replica exchange with solute tempering (REST2). *J Phys Chem B* **115**, 9431-9438, (2011). **PMC3172817**
129. Jorgensen, W., Maxwell, D. & Tirado-Rives, J. Development and testing of the OPLS all-atom force field on conformational energetics and properties of organic liquids. *J. Am. Chem. Soc.* **118**, 11225–11236, (1996).
130. Halgren, T. A. Merck molecular force field .1. Basis, form, scope, parameterization, and performance of MMFF94. *J Comput Chem* **17**, 490-519, (1996).
131. Wang, J., Wolf, R. M., Caldwell, J. W., Kollman, P. A. & Case, D. A. Development and testing of a general amber force field. *J. Comp. Chem.* **25**, 1157-1174, (2004).
132. Vanommeslaeghe, K., Hatcher, E., Acharya, C., Kundu, S., Zhong, S., Shim, J., Darian, E., Guvench, O., Lopes, P., Vorobyov, I. & Mackerell, A. D., Jr. CHARMM general force field: A force field for drug-like molecules compatible with the CHARMM all-atom additive biological force fields. *J. Comp. Chem.* **31**, 671-690, (2010).
133. Jorgensen, W. L. & Tirado-Rives, J. Molecular modeling of organic and biomolecular systems using BOSS and MCPRO. *J. Comp. Chem.* **26** 1689–1700, (2005).
134. Zhang, C. H., Stone, E. A., Deshmukh, M., Ippolito, J. A., Ghahremanpour, M. M., Tirado-Rives, J., Spasov, K. A., Zhang, S., Takeo, Y., Kudalkar, S. N., Liang, Z., Isaacs, F., Lindenbach, B., Miller, S. J., Anderson, K. S. & Jorgensen, W. L. Potent Noncovalent Inhibitors of the Main Protease of SARS-CoV-2 from Molecular Sculpting of the Drug Perampanel Guided by Free Energy Perturbation Calculations. *ACS Cent Sci* **7**, 467-475, (2021).
135. Herrmann, U. S., Schutz, A. K., Shirani, H., Huang, D., Saban, D., Nuvolone, M., Li, B., Ballmer, B., Aslund, A. K., Mason, J. J., Rushing, E., Budka, H., Nystrom, S., Hammarstrom, P., Bockmann, A., Caflisch, A., Meier, B. H., Nilsson, K. P., Hornemann, S. & Aguzzi, A. Structure-based drug design identifies polythiophenes as antiprion compounds. *Sci Transl Med* **7**, 299ra123, (2015).

# Endnotes

[i] The now common expressions "computer alchemy" or "alchemical free-energy perturbation" began to appear in the literature around 1985 (Berendsen,[36,37] Fleishman and Brooks, and Karplus and co-workers[39,40]).

[ii] Natural thermodynamic integration (TI), which traces changes in thermodynamic properties by varying macroscopic state variables (e.g., pressure or temperature), is conceptually distinct from Kirkwood's formulation. In natural TI, the integration proceeds along physically realizable equilibrium paths, with fixed molecular interactions, making it particularly useful for studying phase behavior or response functions. Kirkwood's TI, by contrast, involves a coupling parameter ($\lambda$) that modulates the Hamiltonian to interpolate between two molecular systems – typically involving changes in identity or connectivity – thus creating non-physical intermediate states. This framework defines what are now called alchemical transformations. Free-energy perturbation (FEP), similarly, can operate in both natural and alchemical contexts. Alchemical FEP modifies the Hamiltonian to compute differences between distinct chemical systems (e.g., ligand A to ligand B), while non-alchemical FEP evaluates perturbations to the same system (e.g., small changes in temperature, pressure, ionic strength), without altering molecular identity.

[iii] In a practical numerical free-energy calculations, it is normally assumed that the thermodynamic coupling parameter $\lambda$ should take chosen fixed values between the end-point states. An interesting idea introduced by Tidor has been to incorporate the evolution of the coupling parameter $\lambda$ as part of the dynamic propagation itself, a method now called "$\lambda$-dynamics".[44,45]

[iv] Historically, the word "perturbation" was included in the title of the 1954 paper. Furthermore, the work was presented as a method "by which the thermodynamic properties of one system may be related to those of a slightly different system". However, FEP as expressed by Eq. (2) is not a perturbative method in the traditional sense, as it does not rely on a series expansion (e.g., Taylor or cumulant expansion) to approximate changes. Instead, it employs an exact expression derived from statistical mechanics, Eq (3) $\Delta A = -k_B T \ln \langle e^{-\Delta U/k_B T} \rangle_0$, which provides the full free-energy difference between a reference potential energy $U_0$ and a target potential energy $U_1$ in a single ensemble average from $U_0$ (as specified by the subscript 0 on the averaging symbol). However, in practice, the method is most effective when the two ensembles supported by $U_1$ and $U_0$ exhibit significant overlap; otherwise, poor convergence occurs. In this operational sense, FEP behaves perturbatively: it yields reliable results primarily when the change between reference and target is sufficiently small. The expression FEP with regards to biomolecular simulations begins to appear in early 1990's.[26] Preceding the publication of the expression by Zwanzig in 1954, thermodynamic perturbation theories were formulated by others. Sir Rudolf E. Peierls (1907-1995) formulated a 2nd order perturbative expansion of the partition function in a study of diamagnetic susceptibility of free electrons [R. Peierls, Zur theorie des diamagnetismus von leitungselektronen, Z. Phys. 1933, 80, 763–791]. Furthermore, in the Russian edition of the encyclopedic Course of Theoretical Physics by Lev Landau (1908-1968) and Evgeny Lifshitz (1915-1985), a section on "Thermodynamic Perturbation Theory" in the fifth volume defines the partition function and its expansion in powers of V to second order citing the previous work by Peierls [L. D. Landau and E. M. Lifshitz, Statisticheskaia fizika (klassicheskaia i kVantovaia), Gostekhizdat, 1951; pp 112-115].

[v] It should be noted that in non-ergodic systems, or in systems characterized by complex potential energy landscapes, ensemble averaging can become problematic. Under such conditions, convergence may be impaired by insufficient phase space exploration, leading to significant errors in free-energy estimates. This issue commonly arises in systems with multiple local minima or large energy barriers, which can trap the dynamics in metastable regions and prevent adequate sampling. To address these limitations, researchers have developed enhanced sampling strategies that improve configurational exploration and accelerate convergence. These include, when appropriate, umbrella sampling,[49,50] replica-exchange methods (later known as parallel tempering),[51.] and metadynamics,[52] each of which expands sampling across otherwise inaccessible regions of phase space.

[vi] Ben Widom, email communication with the authors, December 4, 2022: "I did not set out (in 1963) to contribute to numerical potential-energy calculations, only to analytical approaches. I must have known of earlier work by Kirkwood and by Zwanzig, but not in direct connection to what I was doing, or I would otherwise have made reference to them in my paper." This remark underscores that the insertion method was developed independently of the formal free-energy perturbation frameworks, even if later retrospectives have emphasized conceptual parallels.

[vii] William Hoover (born 1936) and Francis Ree (1936–2020)[56] pioneered a computational strategy for determining entropy differences across solid–fluid phase transitions by stabilizing crystalline configurations through an artificial external field. Their approach – employing a single-occupancy constraint within a MC framework – enabled the construction of reversible thermodynamic paths for systems where direct observation of coexistence was hindered by finite-size effects and metastability. Although not framed as a free-energy calculation per se, their method laid significant conceptual groundwork for constrained sampling strategies, and indirectly informed later work, such as the 1969 contribution by Hansen and Verlet, who adapted similar ideas to study phase equilibria in Lennard-Jones systems using homogeneous sampling constraints.

[viii] As we have seen, Valleau's group had previously introduced multistage sampling[16] and importance sampling[17] to improve the efficiency of free-energy calculations in MC simulations. These techniques targeted low-probability regions of configuration space – critical for accurate entropy and free-energy estimates – by deploying either sequential bridging distributions (in multistage sampling) or biased probability functions (in importance sampling). Multistage sampling accomplished this through a series of overlapping ensemble distributions, each providing incremental overlap with its

neighbors, thereby enabling controlled traversal of high-energy or poorly sampled regions. Importance sampling, in turn, reweighted the Boltzmann distribution itself to emphasize rare but thermodynamically significant configurations, effectively concentrating computational effort where it mattered most. Umbrella sampling, by contrast, offered a unified framework for achieving similar goals using a single, continuously biased distribution. Instead of stitching together multiple sampling windows or applying global reweighting, it introduced a biasing potential that reshaped the energy landscape, flattening barriers and improving the accessibility of otherwise inaccessible states. In facilitating continuous sampling across critical regions and enabling efficient reweighting a posteriori to recover unbiased observables, umbrella sampling eliminated the need for successive sampling stages. This conceptual shift – from stagewise interpolation to local energy modification – made the method particularly attractive for systems with sharp phase transitions or discontinuous energy landscapes; it marked a methodological consolidation that combined statistical precision with computational tractability in a single sampling step.

[ix] Arieh Warshel, Zoom conversation with the authors, May 27, 2024.

[x] It is noteworthy that McCammon's insight occurred during the 1983 CECAM (*Centre Européen de Calcul Atomique et Moléculaire*) meeting in Orsay. Founded in 1969 by Carl Moser (1922–2004) in Paris, CECAM was initially located on the Orsay campus, relocated to Lyon in 1994, and moved to Lausanne in 2009. From its inception, CECAM served as a crucial interdisciplinary hub while fostering collaboration among researchers in physics, chemistry, and biology, and catalyzing conceptual breakthroughs in molecular simulation.[72] Notably, the 1977 MD simulation of a protein was carried out during a CECAM workshop in Orsay.[20]

[xi] The publication of this article encountered several difficulties, as explained by Andrew McCammon (email communication with the authors, May 3, 2025): "The manuscript was first submitted to Nature in 1983, but they didn't find it interesting. Then it was submitted to Chemical Physics Letters and was also rejected. It finally found a home in the little-noticed journal, Computers & Chemistry (which currently has the first author's name misspelled as Tembre online)."

[xii] William Jorgensen, Zoom conversation with the authors, July 10, 2024, and email communication, August 30, 2024. It was revealed that the custom code Jorgensen developed in 1985 evolved from an earlier MC program he had originally written in 1978 for simulating simple liquids in the NPT ensemble. This program, later known as BOSS (Biochemical and Organic Simulation System), was extended to handle alchemical transformations through "double-wide sampling" with windowing, which enabled precise on-the-fly averaging of configurations. At the time, this functionality was not available in other platforms, such as GROMOS, which used a single-step FEP approach. Jorgensen also noted that AMBER, although available in 1985, was optimized for biomolecular systems and employed NVE rather than NPT ensemble dynamics, which limited its direct applicability to solvation free-energy studies. He remarked that early FEP calculations using AMBER were prone to numerical instabilities and inadequate convergence due to the lack of ensemble control and insufficient configurational overlap between the initial and final states, particularly in systems with strong electrostatic perturbations or flexible solutes.

[xiii] William Jorgensen, Zoom conversation with the authors, July 10, 2024. Jorgensen explained that his use of free-energy perturbation drew directly from Donald A. McQuarrie's 1976 textbook[73] on statistical mechanics. Jorgensen recalled chapter 14 on "Perturbation Theory of Liquids" as being particularly formative. This is what inspired him to do free energy calculations for reactions in solution. The chapter begins with Zwanzig's 1954 "Statistical Mechanical Perturbation Theory", which why Jorgensen referred to this approach as "statistical perturbation theory" in his early papers. The common name "FEP" was adopted later. McQuarrie's exposition also presented, in detail, the Barker–Henderson and Weeks–Chandler–Andersen perturbation theories of liquids— approaches that operationalized Zwanzig's formalism for realistic intermolecular potentials, circulated widely among statistical mechanicians active at the interface with computations, and were well established in the liquid-state literature before their adaptation to biomolecular contexts. In this context, McQuarrie's textbook operated less as a reference manual than as a framework for building computational methods from first principles. The link between these liquid-state perturbation frameworks and their later adaptation to molecular mechanics Hamiltonians illustrates how pedagogical literature could transmit advanced theoretical constructs into the working repertoire of researchers in molecular simulation, at a stage when the boundaries between theoretical chemistry, statistical mechanics, algorithmic design, and code implementation were markedly more permeable than in later decades.

[xiv] Piotr Cieplak, email communication with the authors, December 2022.

[xv] David Pearlman, Zoom conversation with the authors, February 25, 2025.

[xvi] William Jorgensen, email communication with the authors, May 2, 2025: "I had given a talk at an NSF workshop that Martin Karplus attended where I had presented the FEP results for the first time. The Workshop on Computer Simulation of Quantum and Classical Systems; April 26-28, l985, Harriman, New York (Chemistry) was organized by Bruce Berne. Martin's letter is dated May 3, 1985, so a week later. I remember that Martin was sitting next to Peter Wolynes during my talk. My sense was that they and others thought my results were interesting but impractical owing to the required computer time."

[xvii] William Jorgensen, email communication with the authors, May 16, 2025: "I started my PhD at Harvard in August 1970. I had visited there in the spring and had a meeting with Martin in his office at that time. I had intended to work for Martin, but by the time I got there he had taken a leave and went to Paris. It was unclear when/if he would come back. He was pretty much gone for two years, as I recall. So, I did not have him for a course; I believe he normally taught undergrads anyway. I had courses from Dudley Herschbach, Roy Gordon and people

in Applied Physics. My Ph.D. advisor was organic chemist Elias James Corey (Born July 12, 1928; Nobel Prize in Chemistry, 1990; currently Sheldon Emory Professor Emeritus at Harvard University), but I was in the Chem Phys program and knew many people in Prince House including Klaus Schulten and Attila Szabo. David Case was in some of my classes. I don't think that I officially met Andy McCammon until he had a job interview at Purdue, where I was from 1975, around 1977. He went to Houston, and I visited him there in 1984. Andy is 2 years older than me but graduated from Harvard after me as I started at Harvard when I was 20 and Andy had a time in the Peace Corps. After Martin returned from Paris, I had multiple interaction with him; some had to do with QM calcs for surfaces that he wanted for gas-phase dynamics calcs. Then we exchanged when I was developing water models; he and Peter Kollman both had preprints of the 1983 TIP3P/TIP4P paper that led to its incorporation in CHARMM and AMBER."

[xviii] Although commonly described as an MD program, CHARMM (Chemistry at HARvard Macromolecular Mechanics) was originally conceived as a general-purpose computational chemistry platform. In addition to MD simulations, the software included capabilities for energy minimization, normal mode analysis, MC sampling, and potential energy surface exploration. The original publication by Brooks et al. introduced this multi-functional design, which provided a modular framework for extending CHARMM's capabilities. Over subsequent decades, the software expanded substantially to incorporate advanced methods such as free-energy perturbation, TI, hybrid QM/MM simulations, and enhanced sampling techniques.[83]

[xix] After working briefly for a few months as a graduate student with Hans Delbrück, Karplus decided to do his PhD with John Kirkwood, possibly on the application of statistical mechanics to charge fluctuations in proteins. He also worked for him as teaching assistant for his *Advanced Thermodynamics* class. However, when Kirkwood moved to Yale in 1951, Karplus decided to stay at Caltech to do his PhD with Linus Pauling. The book "*Chemical Thermodynamics*" by Kirkwood and Oppenheim (McGraw-Hill, 1961) is based on the lecture notes of Martin Karplus, Alex Rich and Irwin Oppenheim for Kirkwood's course at Caltech.[85]

[xx] Jiali Gao, Zoom conversation with the authors, March 20, 2025.

[xxi] Wilfred van Gunsteren, Zoom conversation with the authors, July 12, 2024; Tjerk P. Straatsma, email communication with the authors, March 11, 2025. GROMOS (GROningen MOlecular Simulation) was originally developed at the University of Groningen in the late 1970s and early 1980s by Herman Berendsen, Wilfred van Gunsteren and collaborators. GROMOS marked the consolidation of a biomolecular force field and simulation program in the European context. Designed primarily for MD simulations of liquids and biomolecules, GROMOS provided early capabilities for modeling solvated systems using empirical force fields. Its first major documentation appeared in 1987.[74] Free-energy methodologies were implemented in GROMOS in the early 1990's on the basis of a single-topology approach similar in spirit to the one used in AMBER, although with differences in technical details.

[xxii] Tjerk Straatsma, email communication with the authors, March 11, 2025. ARGOS (Alchemical free-energy Rigorous General-purpose Optimization System), developed by Tjerk P. Straatsma beginning in 1984 and later extended with J. Andrew McCammon, was conceived as a flexible, vectorized MD platform focused on free-energy perturbation and thermodynamic integration methodologies.[88] ARGOS's modular structure later contributed to the molecular dynamics components of NWChem (Northwest Chemistry, named after the Pacific Northwest National Laboratory, PNNL), a large-scale computational chemistry software suite initiated at Pacific Northwest National Laboratory around 1995.[89] Designed for massively parallel architectures, NWChem integrates quantum chemistry, molecular dynamics, and free-energy methods to enable large-scale simulations across chemical, biological, and material science domains. After Straatsma moved to Houston to join McCammon's group as a postdoctoral researcher, ARGOS was adopted by other research groups, including Martin Zacharias (born 1961) at Jacobs University Bremen (now at TUM Munich), Rebecca Wade (born 1964) at Heidelberg University, and Volkhard Helms (born 1966) at Saarland University. The program was formally described in 1990 in a publication by Straatsma and McCammon. Developed for efficient vectorization and modular extensibility, ARGOS supported both free-energy computations and broader MD applications. Its computational framework later contributed to the MD capabilities implemented in NWChem (Northwest Chemistry) after 1995.

[xxiii] A CECAM meeting held in September 1987 at Orsay, and a conference sponsored by Alliant at the John von Neumann Supercomputer Center in Princeton—attended by one of the present authors (BR)—reflected the momentum surrounding these developments.

[xxiv] David Case, discussion with one of the authors during the ACS meeting in Indianapolis, March 2023.

[xxv] David Pearlman, Zoom conversation with the authors, February 25, 2025. Pearlman emphasized that the use of a single-topology framework and the neglect of explicit treatment of dummy particles were understood at the time as practical approximations for small side-chain transformations, but acknowledged that these strategies would likely be insufficient for larger, more complex side chains such as tryptophan (Trp), phenylalanine (Phe), and tyrosine (Tyr).

[xxvi] Wilfred van Gunsteren, Zoom conversation with the authors, July 12, 2024. Early on, there was a lot of interest in representing the total free energy between two endpoints as a sum of separate chemically identifiable components. This led to several arguments and counterarguments about the validity of such decomposition and analysis.[91,92,102]

[xxvii] Stefan Boresch, Zoom conversation with the authors, March 11, 2025.

[xxviii] Wade et al. (1991) used the TIP3P water model. Although TIP4P offers improved accuracy for certain physical properties of water, TIP3P was more commonly used at the time due to its lower computational demands and widespread adoption in biomolecular simulations. In Wade and coworkers' study, TIP3P provided a practical balance between computational cost and sufficient accuracy for modeling hydration dynamics in

protein cavities, where local interactions were of primary concern.

[xxix] In 1989, Jorgensen avoided alchemical decoupling in a problem of molecular association, and instead relied on a potential of mean force (PMF) formulation of the binding constant.[71] However, this PMF formulation, which is formally correct, was not immediately adopted in the following calculations. In 2005, Woo and Roux extended the PMF formulation of molecular association to account for the effects of conformational, orientational and positional restraints.[115]

[xxx] Mike Gilson, Zoom conversation with the authors, March 17, 2025.

[xxxi] Rich Friesner, Zoom conversation with the authors, March 21, 2025.

[xxxii] Alchemical ABFE are particularly important because they rigorously account for complex thermodynamic contributions such as solvation, conformational flexibility, and entropy. Scoring approaches, such as Linear Interaction Energy (LIE)[122] and Molecular Mechanics-Poisson-Boltzmann Surface Area (MM-PBSA) are endpoint approximations, often of unknown validity.[123] While they can be useful for high-throughput screening, they generally offer less precision ABFE.

[xxxiii] Despite the recognition granted to ABFE nowadays, the pioneering formal developments were often misunderstood and perceived negatively by members of the community. The Roux et al 1996 article[116] submitted to the *Journal of Molecular Biology* in September 1995 was rejected in January 1996 after a long delay. It was then submitted to *Biophysical Journal* in March 1996 and finally published in August 1996. It is interesting to note that McCammon, co-author of Gilson et al 1997[117], was the handling editor of the submission. The Gilson 1997 article was submitted to *Biophysical Journal* in May 1996 but published only in March 1997. Roux, who was the handling editor of the submission, had to override a negative reviewer recommending rejection who kept insisting "There is nothing new here that isn't already in a standard statistical mechanics textbook." Roux also recalls presenting this work at the ACS meeting in Orlando in August 1996 during a session entitled "Molecular Dynamics and Free-Energy Perturbations Free-Energy Perturbation", where an unimpressed member of the audience asked "Isn't all this already in Hill's textbook?" In the same session, Jans Herman was presenting the work of Wang and Hermans (1997), and Bruce Bush (co-author of Gilson et al 1997) was presenting free-energy simulation with the Merck force field.

[xxxiv] Andrew J. McCammon, email communication with the authors, May 4, 2025.

[xxxv] Alliant Computer Systems introduced the FX/8 minisupercomputer in the mid-1980s. It was a shared-memory multiprocessor adopted in research computing centers for scientific workloads and combined multiple computational units with interactive processors and vector capabilities. Technical descriptions and performance results are reported in W. Abu-Sufah and A. D. Malony, Vector Processing on the Alliant FX/8 Multiprocessor, in Proceedings of the International Conference on Parallel Processing (IEEE Computer Society Press, Los Alamitos, 1986), pp. 559–566, and in R. T. Dimpsey, Performance Analysis of the Alliant FX/8 Multiprocessor Using Statistical Clustering, University of Illinois at Urbana-Champaign Technical Report UILU-ENG-88-2255 (CSG-91), 1988. Hardware specifications are also preserved in the Computer History Museum's catalog entry for the FX/8 CPU.

[xxxvi] One notable workshop held at the John von Neumann supercomputing center of Princeton University in 1989 included, Martin Karplus, Wilfred van Gunsteren, William Jorgensen, Charles Brooks, Monte Pettitt, Lennard Nilsson, and many more investigators.

[xxxvii] Andrew J. McCammon, email communication with the authors, May 2, 2025.

[xxxviii] The term first appeared in 1984 as the topic of a public debate at the annual meeting of AAAI (then called the "American Association of Artificial Intelligence"). Roger Schank and Marvin Minsky – two leading AI researchers who experienced the "winter" of the 1970s–warned the business community that enthusiasm for AI had spiraled out of control in the 1980s and that disappointment would certainly follow. They described a chain reaction, similar to a "nuclear winter", that would begin with pessimism in the AI community, followed by pessimism in the press, followed by a severe cutback in funding, followed by the end of serious research. Three years later the billion-dollar AI industry began to collapse. There were two major winters approximately 1974–1980 and 1987–2000, and several smaller episodes. Enthusiasm and optimism about AI has generally increased since its low point in the early 1990s. Beginning about 2012, interest in artificial intelligence (and especially the sub-field of machine learning) from the research and corporate communities led to a dramatic increase in funding and investment, leading to the current (as of 2025) AI boom.

[xxxix] Frank Noe, email communication, April 1, 2025.